\documentclass[10pt]{article}
\usepackage{epsfig,amsmath,amsthm,graphicx,amssymb,overpic}
\usepackage{epsfig,amsmath,graphicx,amssymb,overpic,cite}
\usepackage{listings,diagbox}

\def\be{\begin{equation}}
\def\ee{\end{equation}}
\def\bee{\begin{eqnarray}}
\def\ene{\end{eqnarray}}
\def\bes{\begin{subequations}}
\def\ees{\end{subequations}}

\def\v{\vspace{0.1in}}

\begin{document}

\baselineskip=13pt
\renewcommand {\thefootnote}{\dag}
\renewcommand {\thefootnote}{\ddag}
\renewcommand {\thefootnote}{ }

\pagestyle{plain}

\begin{center}
\baselineskip=16pt \leftline{} \vspace{-.3in} {\Large \bf Data-driven peakon and periodic peakon travelling wave solutions of some nonlinear dispersive equations via deep learning} \\[0.2in]
\end{center}

\begin{center}
Li Wang and Zhenya Yan$^{*}$\footnote{$^{*}${\it Email address}: zyyan@mmrc.iss.ac.cn (Corresponding author)}  \\[0.03in]
{\it Key Laboratory of Mathematics Mechanization, Academy of Mathematics and Systems Science, \\ Chinese Academy of Sciences, Beijing 100190, China \\
 School of Mathematical Sciences, University of Chinese Academy of Sciences, Beijing 100049, China} \\
\end{center}

\vspace{0.3in}

{\baselineskip=13pt


\vspace{-0.28in}
\begin{abstract}
 In the field of mathematical physics, there exist many physically interesting nonlinear dispersive equations with peakon solutions, which are solitary waves with discontinuous first-order derivative at the wave peak. In this paper, we apply the multi-layer physics-informed neural networks (PINNs) deep learning  to successfully study the data-driven peakon and periodic peakon solutions of some well-known nonlinear dispersion equations with initial-boundary value conditions such as the Camassa-Holm (CH) equation, Degasperis-Procesi equation, modified CH equation with cubic nonlinearity, Novikov equation with cubic nonlinearity, mCH-Novikov equation, $b$-family equation with quartic nonlinearity,
 generalized modified CH equation with quintic nonlinearity, and etc. These results will be useful to further study the peakon solutions and corresponding experimental design of nonlinear dispersive equations.

\vspace{0.1in}
\end{abstract}

{\it Key words:} Nonlinear dispersive equation; initial-boundary value conditions; physics-informed neural networks; deep learning; data-driven

\baselineskip=13pt

\section{Introduction}

The well-known nonlinear dispersive Camassa-Holm (CH) equation can be used to describe the propagation of shallow water waves, and
deduced by using the distinct approaches such as the symmetry method~\cite{ch1}, an asymptotic expansion directly in the Hamiltonian for Euler's equations~\cite{ch}, and the tri-Hamiltonian duality related to the  Korteweg-de Vries equation~\cite{tri}. The CH equation admitted the peaked travelling wave solutions (alias {\it peakons}, i.e., solitary waves with discontinuous first-order derivative at the wave peak)~\cite{ch}. After that,  much attention has been paid to some physically interesting nonlinear wave equations describing breaking waves and possessing the peakon solution in the form $u_t-u_{xxt}+F(u,u_x,u_{xx}, u_{xxx})=0$ such as the Degasperis-Procesi (DP) equation~\cite{dp}, $b$-family equation~\cite{bch1,bch2}, Fokas-Olver-Rosenau-Qiao (FORQ) equation (alias  modified CH (mCH) equation)~\cite{mch1,mch3,tri}, mCH-CH equation~\cite{mch-ch}, Novikov equation~\cite{nov}, mCH-Novikov equation~\cite{mch-n}, and etc. Up to now, it is still an interesting subject to study these nonlinear peakon equations via distinct approaches.

Recently, mechanics learning with deep neural networks plays a more and more important role
in many fields~\cite{ml1,ml2}. Particularly, various deep neural network learning approaches~\cite{dl-pde1,dl-pde2,dl-pde3,dl-pde4,dl-pde5,dl-pde6,Han,raiss18,raiss19} have been paid more and more attention in the study of linear and nonlinear partial differential equations (PDEs). The physics-informed neural network (PINN) approach~\cite{raiss18,raiss19} were presented, and has been applied to many linear and nonlinear PDEs~\cite{raiss20,pinn1,pde20,yan20,yan21}, as well as other types of equations such as the fractional PDEs~\cite{fpde}, and stochastic differential equations~\cite{spde}. More recently,  Shin, {\it et al}~\cite{proof-pinn} theoretically showed the consistency of PINNs for the linear second-order elliptic and parabolic type PDEs.

In this paper, we would like to use the PINN deep learning to consider the Cauchy problem of the nonlinear dispersive equations with peakon solutions
\bee\label{pde}
\left\{\begin{array}{l}
(1-\partial_x^2)u_t+{\cal N}(u,u_x,u_{xx},u_{xxx},...)=0,\quad x\in (L_1, L_2),\quad t\in (0, T), \vspace{0.1in}\\
u(x,0)=u_0(x), \quad  x\in [L_1, L_2], \vspace{0.1in}\\
u(L_1, t)=u(L_2, t),\quad t\in [0, T],
\end{array}\right.
\ene
where the subscripts denote the partial derivatives (e.g., $u_t=\partial u/\partial t$), $\partial_x=\partial/\partial x$, ${\cal N}(u,u_x,u_{xx},u_{xxx},...)$ is some nonlinear function of related variables such that Eq.~(\ref{pde}) contains many famous
nonlinear peakon equations, for example:
\begin{itemize}

\item [i)] The generalized $b$-family (gbf) equation with nonlinearities of degree $(k+1)$~\cite{gbf,gbf2,gbf3,gbf4}
\bee
 \label{gbf}
(1-\partial_x^2)u_t+u^k\omega_x+bu^{k-1}u_x\omega=0,\quad \omega:=u-u_{xx},\,\,\, k\in \mathbb{N}^+,\quad b\in \mathbb{R},
\ene
which contains the well-known models: ia) the integrable CH equation~\cite{ch,ch1} for $k=1, b=2$
\bee \label{ch}
(1-\partial_x^2)u_t+3uu_x-2u_xu_{xx}-uu_{xxx}=0
\ene
admitting the second-order Lax pair~\cite{ch};
ib) the integrable DP equation~\cite{dp} for $k=1, b=3$
 \bee \label{dp}
(1-\partial_x^2)u_t+4uu_x-3u_xu_{xx}-uu_{xxx}=0,
\ene
which is related to a negative flow of the Kaup-Kupershmidt hierarchy by a reciprocal transformation, and admits the third-order Lax pair~\cite{bch1};
ic) the $b$-family equation~\cite{bch1,bch2} for $k=1$
 \bee \label{bf}
(1-\partial_x^2)u_t+(b+1)uu_x-bu_xu_{xx}-uu_{xxx}=0,
\ene
id) the Novikov equation with cubic nonlinearity~\cite{nov} for $k=2,\, b=3$
\bee \label{nov}
(1-\partial_x^2)u_t+4u^2u_x-3uu_xu_{xx}-u^2u_{xxx}=0,
\ene
which is also completely integrable, and related to a negative flow in the Sawada-Kotera hierarchy by a reciprocal transformation, as well as admits the third-order Lax pair~\cite{nov-p}; and ie) the modified Novikov equation~\cite{mnov} for $k=2$
\bee \label{novb}
(1-\partial_x^2)u_t+(b+1)u^2u_x-buu_xu_{xx}-u^2u_{xxx}=0;
\ene

\item [ii)] The generalized modified Camassa-Holm (gmCH) equation with nonlinearities of degree $(2k+1)$~\cite{gmch1,gmch2}
\bee \label{gmch}
(1-\partial_x^2)u_t+[(u^2-u_x^2)^k(u-u_{xx})]_x=0,\quad k\in \mathbb{N}^+,
\ene
containing the integrable FORQ equation~\cite{mch1,tri,mch3} (or the mCH equation) with cubic nonlinearity for $k=1$
\bee \label{mch}
(1-\partial_x^2)u_t+[(u^2-u_x^2)(u-u_{xx})]_x=0,
\ene
which is also completely integrable, and admits the second-order Lax pair~\cite{mch3}.

\item [iii)] The generalized mCH and $b$-family equation with nonlinearities of degree ${\rm max}\{2k+1, s+1\}$
\bee \label{gmchb}
(1-\partial_x^2)u_t+k_1[(u^2-u_x^2)^k\omega]_x+k_2(u^s\omega_x+bu^{s-1}u_x\omega)=0,\quad k,\, s\in \mathbb{N}^+,
\ene
containing iiia) the mCH-CH equation for $k=s=1,\, b=2$~\cite{mch-ch}
\bee \label{mch2}
(1-\partial_x^2)u_t+k_1[(u^2-u_x^2)\omega]_x+k_2(u\omega_x+2u_x\omega)=0,\quad k_1,\,k_2\in \mathbb{R},
\ene
and iiib) the mCH-Novikov equation with cubic nonlinearity~\cite{mch-n}
\bee \label{mchn}
(1-\partial_x^2)u_t+k_1[(u^2-u_x^2)\omega]_x+k_2(u^2\omega_x+3uu_x\omega)=0,\quad k_1,\,k_2\in \mathbb{R},
\ene

\item [iv)] The combination of generalized mCH and $b$-family equations
\bee\label{gpe1}
(1-\partial_x^2)u_t+\sum_{k=1}^Nk_j[(u^2-u_x^2)^k\omega]_x+\sum_{s=1}^Mk_s(u^s\omega_x+bu^{s-1}u_x\omega)=0,
\ene

\item [v)] The general family of nonlinear peakon equations~\cite{gmch2}
 \bee\label{gpe2}
 (1-\partial_x^2)u_t+f_1(u,u_x)\omega+(f_2(u, u_x)\omega)_x=0,
 \ene
where $f_j$'s are some functions of $u$ nd $u_x$.

\end{itemize}

Notice that i) The generalized $b$-family equation (\ref{gbf}) containing the CH equation (\ref{ch}), DP equation (\ref{dp}), $b$-family equation (\ref{bf}), and Novikov equation (\ref{nov}), possesses the peakon solution $u(x,t)=c^{1/k}e^{-|x-ct|}$, and periodic peakon solution $u(x,t)=c^{1/k}\,{\rm sech} \pi\, p(x)$ with $p(x)=\cosh(x-ct-2\pi[(x-ct)/(2\pi)]-\pi)$ and $\lfloor\cdot\rfloor$ denoting the floor function or the greatest integer function~\cite{gbf,gbf2,gbf3,gbf4}; ii) The mCH equation (\ref{mch}) has the peakon solution~\cite{mch-p} $u(x,t)=\sqrt{3c/2}e^{-|x-ct|}, \, c>0$ and periodic peakon solution~\cite{mch-pp} $u(x,t)=\sqrt{3c/(2\cosh^2\pi+1)}\,p(x), \, c>0$; iii) The mCH-Novikov equation (\ref{mchn}) has the peakon solution~\cite{mch-n} $u(x,t)=\sqrt{3c/(2k_1+3k_2)}e^{-|x-ct|}$ and periodic peakon solution $u(x,t)=\sqrt{3c/[k_1(2\cosh^2\pi+1)+3k_2\cosh^2\pi]}\,p(x);$
iv) The generalized mCH equation (\ref{gmch}) with $k=2$  has the peakon solution~\cite{gmch-s} $u(x,t)=\sqrt[4]{15c/8}e^{-|x-ct|}$ and
periodic peakon solution $u(x,t)=\sqrt[4]{15c/[\sinh\pi(8\cosh^4\pi+4\cosh^2\pi+3)]}\,p(x)$.

In this paper, we would like to consider the data-driven peakon and periodic peak solutions of the above-mentioned some physical interesting nonlinear PDEs via deep learning. The rest of this paper is arranged as follows: In Sec. 2, we introduce the PINN deep leaning approach for the Cauchy problem of nonlinear PDEs (\ref{pde}). In Sec. 3, we use the PINN deep learning scheme to investigate the data-driven peakon and periodic peakon solutions of some famous nonlinear PDEs such as the CH equation (\ref{ch}), DP equation (\ref{dp}), Novikov equation (\ref{nov}), $b$-family equation (\ref{gbf}) with $b=k=3$, mCH equation (\ref{mch}), generalized mCH equation (\ref{gmch}) with $k=2$,  mCH-CH equation (\ref{mch2}), and the mCH-Novikov equation (\ref{mchn}). In Sec. 4, we give some conclusions and discussions.

\section{The PINN deep learning scheme of Eq.~(\ref{pde})}


For the given Cauchy problem (\ref{pde}) we consider the residual PINN $f(x,t)$ as
\bee\label{f}
 f(x,t):=(1-\partial_x^2)\widehat u_t+{\cal N}(\widehat u,\widehat u_x,\widehat u_{xx}, \widehat u_{xxx},...),
\ene
where ${\cal N}(\widehat u,\widehat u_x,\widehat u_{xx}, \widehat u_{xxx},...)$ can be chosen as the some physical models mentioned in Sec. 1, a deep neural network $\widehat u(x, t; w, b)$ denotes the continuous latent function with two families of network parameters: the weights $w$ and biases $b$, and can be used to approximate $u(x,t)$ such that one can consider its partial derivatives of arbitrary order with respect to its input with the aid of automatic differentiation (e.g., Tensorflow )~\cite{auto1,auto2}, which differs from the numerical or symbolic differentiation. And then the residual PINN $f(x,t)$ given by Eq.~(\ref{f}) can also be found.

Therefore, with the aid of L-BFGS optimization method~\cite{Liu}, the common parameters in the latent function $\widehat u(x, t)$ and residual PINN $f(x,t)$ can be trained by using the multi-hidden-layer deep NN with some neurons per layer combined with a hyperbolic tangent activation function
and minimizing the whole mean squared error (MSE) loss in the form
\begin{equation} \label{mse}
MSE = MSE_{ie} +  MSE_{b} + MSE_{f},
\end{equation}%
where
\begin{align}
MSE_{ie} &= \frac{1}{N_{int}} \sum_{j=1}^{N_{int}}\left| \widehat u(x_{0}^{j}, 0)- u(x_{0}^{j}, 0)\right|^{2}
       + \frac{1}{N_{end}} \sum_{j=1}^{N_{end}}\left|\widehat u(x_{T}^{j},  T)- u(x_{T}^{j}, T)\right|^{2},  \label{int_loss} \\
MSE_{b} &=   \frac{1}{N_{b}} \sum_{j=1}^{N_{b}}\left(\left|\widehat u(L_1,t_{b}^{j})-u(L_1,t_{b}^{j})\right|^{2}
       + \left|\widehat u(L_2, t_{b}^{j})- u(L_2, t_{b}^{j})\right|^{2}+
       \left| \widehat u(L_1, t_{b}^{j}) - \widehat u(L_2, t_{b}^{j}) \right|^{2}\right), \label{b_loss}   \\
MSE_{f} &= \frac{1}{N_{f}} \sum_{j=1}^{N_{f}} \left|  f( x_{f}^{j}, t_{f}^{j}) \right|^{2}, \label{f_loss}
\end{align}
the observed measurements $\{\widehat u(x_{0}^j, 0)\}_{1}^{N_{int}}$ and $\{\widehat u(x_{T}^j, T)\}_{1}^{N_{end}}$ of the hidden field $\widehat u(x,t)$ are linked with the sampled initial and end training data $\{x_{0}^j,\, u(x_0^{j}, 0)\}_{1}^{N_{int}}$ and
$\{x_{T}^j,\, u(x_T^{j}, T)\}_{1}^{N_{end}}$, respectively. $\{ \widehat u(L_{1,2}, t_{b}^j) \}_{1}^{N_b}$ connect the selected boundary training data $\{t_b^j,\, u(L_{1,2}, t_{b}^j \}_{1}^{N_b}$, and $\{ x_{f}^j, t_{f}^j \}^{N_f}_{1}$ are connected with the marked points for the PINN $f(x,t)$. As a result, for the randomly chosen points, $MSE_{ie}$, and $MSE_{b}$ represent the  MSE losses of initial-end and periodic boundary data, respectively, and $MSE_{f}$ is associated with the MSE loss of the PINN (\ref{f}). The aims of $MSE_{ie}$ and $MSE_{b}$ try to match the learning solution to exact one for the initial and end data, and boundary data, respectively. The aim of $MSE_f$ is to make the hidden $\widehat u(x,t)$ satisfy the considered physical equation (\ref{pde}).

In what follows we would like to use the PINN deep learning scheme to investigate the data-driven peakon and periodic peakon solutions of some famous nonlinear PDEs such as the CH equation (\ref{ch}), DP equation (\ref{dp}), Novikov equation (\ref{nov}), $b$-family equation (\ref{gbf}) with $b=k=3$, mCH equation (\ref{mch}), generalized mCH equation (\ref{gmch}) with $k=2$,  mCH-CH equation (\ref{mch2}), and the mCHN equation (\ref{mchn}).

\section{The data-driven peakon and periodic peakon solutions}

\subsection{The CH equation}

In the subsection, we use the above-mentioned PINN deep learning approach to consider the data-driven peakon and periodic peakon solutions of the initial-boundary value problem of the CH equation given by Eqs.~(\ref{pde}) and (\ref{ch}).

\v {\it Case 1.} For the peakon solution~\cite{ch} $u(x,t)=ce^{-|x-ct|}$ of the CH equation (\ref{ch}), it is easy to see that when $c>0$, the peakon solution is a {\it right-going} travelling wave solution, and bright peakon solution, whereas $c<0$, it is a  {\it left-going} travelling wave solution, and dark peakon solution. We here consider the stationary peakon solution at $t=0$ as the initial condition
\bee \label{ch-int}
u(x,0)=ce^{-|x|},\quad  x\in [-L, L],
\ene
and the periodic boundary condition $u(-L, t)=u(L, t)$. The considered  residual PINN $f_{ch}(x,t)$ is written as
\bee
 f_{ch}(x,t):=u_t-u_{xxt}+3uu_x-2u_xu_{xx}-uu_{xxx}.
\ene

Here, the hidden neural network $\widehat u(x,t)$ in Python can be defined as
\begin{lstlisting}
def u(x, t):
    u = neural_net(tf.concat([x,t],1), weights, biases)
    return u
\end{lstlisting}
such that the residual PINN $f_{ch}(x,t)$ in Python is written as
\begin{lstlisting}
 def f_ch(x, t):
     u = u(x, t)
     u_t = tf.gradients(u, t)[0]
     u_x = tf.gradients(u, x)[0]
     u_xx = tf.gradients(u_x, x)[0]
     u_xxt = tf.gradients(u_xx, t)[0]
     u_xxx = tf.gradients(u_xx, x)[0]
     f_ch = u_t - u_xxt + 3*u*u_x - 2*u_x*u_xx - u*u_xxx
     return f_ch
\end{lstlisting}

i) For the case $c>0$ in the initial condition (\ref{ch-int}), i.e., the bright peakon solution, we take $c=0.8$ and use the Fourier pseudo-spectral method~\cite{tref} (i.e., one can take Fourier transform in space, and choose the explicit fourth-order Runge-Kutta method in time) to simulate the CH equation (\ref{ch}) with the initial value condition (\ref{ch-int}) and periodic boundary condition. The spatial region is $x\in [-6,\, 6]$ with 512 Fourier modes, and the temporal region is $t\in [0, 2]$ with time-step $\Delta t = 0.004$. As a consequence, we produce the data-set for the PINN pertaining to the CH equation. The training data-set used in  the PINN $f_{ch}(x,t)$ consists of randomly chosen $N_{int}=10$ points from the initial data $u(x,0)$, $N_{end}=10$ points from the end data $u(x,2)$,  $N_b=10$ points pertaining to the periodic boundary data, and $N_f=1000$ points from the spatial-temporal solution zone.

Figures~\ref{ch-bd}(a1) and (c1) exhibit the two-dimensional (2D) and three-dimensional (3D) profiles of the latent bright-peakon solution $\widehat u (x,t)$ learned by a 5-hidden-layer deep PINN $f_{ch}(x,t)$ with 20 neurons per layer combined with a hyperbolic tangent activation function, and  minimizing the MSE loss (\ref{mse}). The comparison of learning solution (red dashed line), numerical solution (blue solid line) and exact solution (green plus) is shown at three different times $t = 0.4,\, 1.0$, and $1.4$ (see Fig.~\ref{ch-bd}(b1)). The $\mathbb{L}_{2}$-norm error between learning solution $\widehat{u}(x,t)$  and numerical solution is 2.73e-02.

\begin{figure}[t]
\begin{center}
\hspace{0.2in} {\scalebox{0.75}[0.75]{\includegraphics{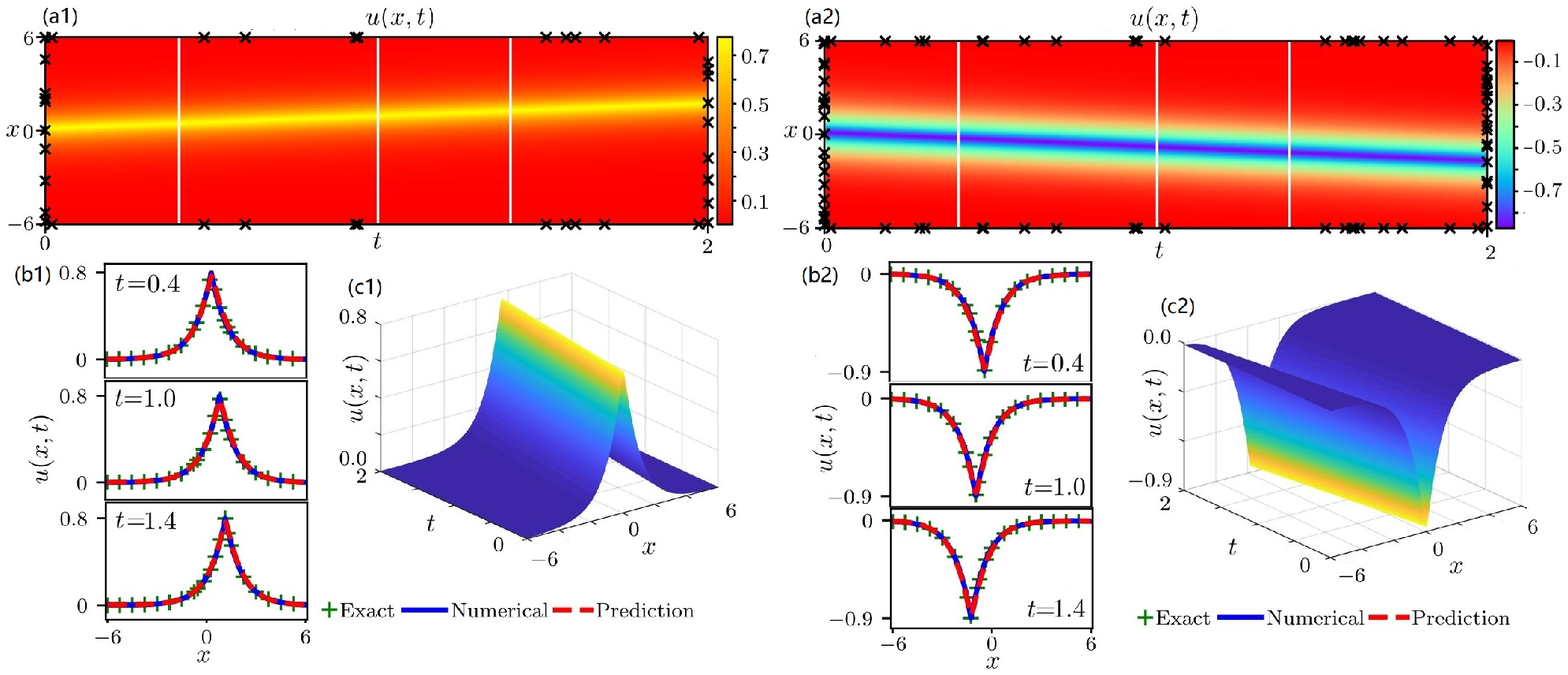}}}
\end{center}
\par
\vspace{-0.06in}
\caption{{\protect\small The CH equation. (a1,a2) The data-driven peakon solutions resulted from the PINN; (b1,b2) The comparisons between the learning, numerical, and exact peakon solutions at the distinct times $t=0.4,\, 1$, and $1.4$; The $\mathbb{L}_{2}$-norm errors $\widehat u(x,t)$ between learning and numerical peakon solutions are (b1) 2.73e-02 and (b2) 4.62e-02; (c1,c2) The 3D profiles of the learning bright and dark peakon solutions. $c=0.8$ for (a1, b1, c1) and $c=-0.9$ for (a2, b2, c2).}}
\label{ch-bd}
\end{figure}
\begin{figure}[!ht]
\begin{center}
\vspace{-0.1in}
{\scalebox{0.45}[0.38]{\includegraphics{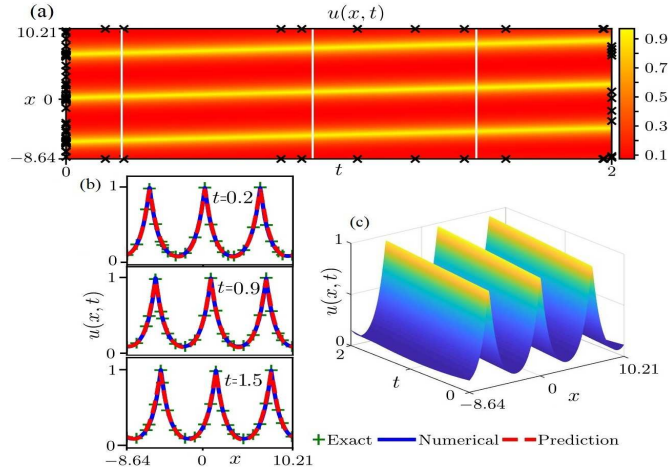}}}
\end{center}
\par
\vspace{-0.1in}
\caption{{\protect\small The CH equation: (a, c) The data-driven periodic peakon solution resulted from the PINN; (b) The comparisons between the learning, numerical, and exact periodic peakon solutions at the distinct times $t = 0.2$, $0.9$, and $1.5$. The $\mathbb{L}_{2}$-norm error between the learning and numerical periodic peakon solutions is  3.56e-02.}}
\label{fch-p}
\end{figure}

ii) For the case $c<0$ in the initial condition (\ref{ch-int}), i.e., dark peakon solution, we take $c=-0.9$ and use the same  pseudo-spectral approach  to produce the data-set for PINN pertaining to the CH equation in the spatio-temporal region $(x,t)\in [-6, 6]\times [0, 2]$ with
the 512 Fourier modes in the $x$ direction and time-step $\Delta t = 0.004$. The training data-set used in  the PINN consists of randomly chosen $N_{int}=20$ points from the initial data $u(x,0)$, $N_{end}=20$ points from the end data $u(x,2)$,  $N_b=20$ points pertaining to the periodic boundary data, and $N_f=1000$ points from the spatio-temporal solution zone. Figures~\ref{ch-bd}(a2) and (c2) exhibit the 2D and 3D profiles of the latent dark-peakon solution $\widehat u (x,t)$ learned by a 6-hidden-layer deep PINN with 20 neurons per layer combined with a hyperbolic tangent activation function. The comparison of learning solution (red dashed line), numerical solution (blue solid line) and exact solution (green plus) is shown at three different times $t = 0.4,\, 1.0$, and $1.4$ (see Fig.~\ref{ch-bd}(b2)).  The $\mathbb{L}_{2}$-norm error between learning solution $\widehat{u}(x,t)$  and numerical solution is 4.62e-02.

\v {\it Case 2.} For the periodic peakon solution~\cite{gbf2} $u(x,t)=c\,{\rm sech} \pi \cosh(x-ct-2\pi\lfloor(x-ct)/(2\pi)\rfloor-\pi)$ of the CH equation (\ref{ch}), we consider the stationary periodic peakon solution at $t=0$ as the initial value condition
\bee\label{ch-p}
u(x,0)=c\,{\rm sech} \pi \cosh\left(x-2\pi\left\lfloor\frac{x}{2\pi}\right\rfloor-\pi\right),\quad x\in [L_1, L_2]
\ene
and the periodic boundary $u(L_1, t)=u(L_2, t)$, as well as $c=1$. Similarly, we use the pseudo-spectral approach to generate
the data-set for the PINN pertaining to the CH equation in the spatio-temporal region $(x,t)\in [-8.64, 10.21]\times [0, 2]$ with
the 512 Fourier modes in the $x$ direction and time-step $\Delta t = 0.004$. The training data-set used in  the PINN consists of randomly chosen $N_{int}=60$ points from the initial data $u(x,0)$, $N_{end}=10$ points from the end data $u(x,2)$,  $N_b=10$ points pertaining to the periodic boundary data, and $N_f=1100$ points from the spatio-temporal solution zone. The hidden solution $\widehat u(x,t)$ is learned by a 6-hidden-layer deep PINN $f_{ch}(x,t)$ with 40 neurons pere layer combined with a hyperbolic tangent activation function.

 Figures~\ref{fch-p}(a) and (c) exhibit the 2D and 3D profiles of the latent periodic peakon solution $\widehat u (x,t)$. The comparison of learning solution (red dashed line), numerical solution (blue solid line) and exact solution (green plus) is shown at three different times $t = 0.2,\, 0.9$, and $1.5$ (see Fig.~\ref{fch-p}(b)). The $\mathbb{L}_{2}$-norm error between learning solution $\widehat{u}(x,t)$  and numerical solution is 3.56e-02.

  Similarly, the data-driven peakon and periodic peakon solutions of the DP equation (\ref{dp}) are also studied in {\bf Appendix A}.

\subsection{The Novikov equation with cubic nonlinearity}

In the subsection, we use the PINN to consider the data-driven peakon solutions of the initial-boundary value problem of the Novikov equation with cubic nonlinearity (\ref{nov}).

\v {\it Case 1.} For the peakon solution~\cite{nov-p} $u(x,t)=\sqrt{c}e^{-|x-ct|}$ with $c>0$ of the Novikov equation (\ref{nov}), which is a {\it right-going} travelling wave solution, and bright peakon solution. It follows from the peakon solutions of CH and Novikov equations that for the
right-going travelling wave with the same velocity, i) when $c>1$, the amplitude of the CH peakon solution is larger than one of the Novikov peakon solution; when $0<c<1$, the amplitude of the CH peakon solution is smaller than one of the Novikov peakon solution; iii) when $c=1$, the amplitudes of the CH and Novikov peakon solutions are same. We here consider the the stationary peakon solution at $t=0$ as the initial condition
\bee \label{nov-int}
u(x,0)=\sqrt{c}\,e^{-|x|},\quad x\in [-L, L]
\ene
and the periodic boundary $u(-L, t)=u(L, t)$. The considered PINN $f_{nov}(x,t)$ is written as
\bee
 f_{nov}(x,t):=u_t-u_{xxt}+4u^2u_x-3uu_xu_{xx}-u^2u_{xxx}.
\ene

We here choose $c=0.36$ in the initial condition (\ref{nov-int}), and use the Fourier pseudo-spectral method (i.e., one can take Fourier transform in space, and choose the explicit fourth-order Runge-Kutta method in time) to simulate the Novikov equation with the initial value condition (\ref{nov-int}) and periodic boundary condition. The spatial region is $x\in [-10,\, 10]$ with 256 Fourier modes, and the temporal region is $t\in [0, 2]$ with time-step $\Delta t = 0.004$. As a result, we produce the data-set for the PINN pertaining to the Novikov equation. The training data-set used in  the PINN $f_{nov}(x,t)$ consists of randomly chosen $N_{int}=20$ points from the initial data $u(x,0)$,
$N_{end}=20$ points from the end data $u(x,2)$,  $N_b=20$ points pertaining to the periodic boundary data, and $N_f=1000$ points
from the spatio-temporal solution zone. Figures~\ref{fnov-bp}(a1) and (c1) exhibit the 2D and 3D profiles of the latent bright peakon solution $\widehat u (x,t)$ learned by a 6-hidden-layer deep PINN $f_{nov}(x,t)$ with 20 neurons per layer combined with a hyperbolic tangent activation function. The comparison of learning solution (red dashed line), numerical solution (blue solid line) and exact solution (green plus) is shown at three different times $t = 0.2, 1.0$ and $1.8$ (see Fig.~\ref{fnov-bp}(b1)). The $\mathbb{L}_{2}$-norm error between learning solution $\widehat{u}(x,t)$  and numerical solution is 3.82e-02.

\begin{figure}[!t]
\begin{center}
\vspace{-0.15in}
\hspace{0.2in} {\scalebox{0.75}[0.75]{\includegraphics{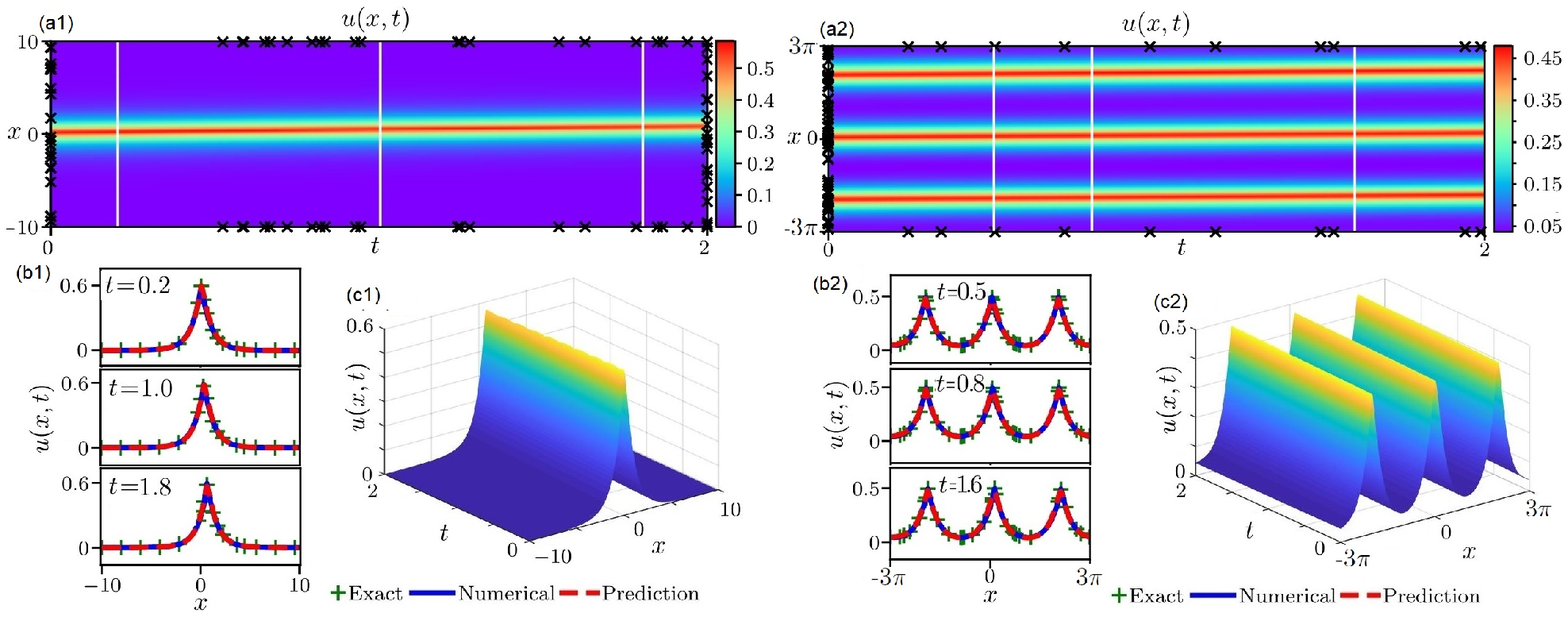}}}
\end{center}
\par
\vspace{-0.05in}
\caption{{\protect\small The Novikov equation. The data-driven peakon (a1,c1) and periodic peakon (a2, c2) solutions resulted from the PINN;
(b1, b2) The comparisons between the learning, numerical, and exact peakon or periodic peakon solutions at the distinct times
(b1) $t=0.2,\, 1$, and $1.8$, or (b2) $t=0.5,\, 0.8$, and $1.6$. The $\mathbb{L}_{2}$-norm errors  between the learning and numerical
solutions are (b2) 3.82e-02 and (b2) 4.56e-02. $c=0.36$ for (a1,b1,c1), and $c=0.25$ for (a2,b2,c2).}}
\label{fnov-bp}
\end{figure}

{\it Case 2.} For the periodic peakon solution~\cite{gbf2,gray} $u(x,t)=\sqrt{c}\,{\rm sech} \pi \cosh(x-ct-2\pi\lfloor(x-ct)/(2\pi)\rfloor-\pi)$ of the Novikov equation (\ref{nov}), we consider the stationary periodic peakon solution at $t=0$ as the initial value condition
\bee\label{nov-p}
u(x,0)=\sqrt{c}\,{\rm sech} \pi \cosh\left(x-2\pi\left\lfloor\frac{x}{2\pi}\right\rfloor-\pi\right),
\ene
and the periodic boundary $u(-L, t)=u(-L, t)$, as well as $c=0.25$. Similarly, we use the pseudo-spectral approach to generate
the data-set for the PINN pertaining to the Novikov  equation in the spatio-temporal region $(x,t)\in [-3\pi, 3\pi]\times [0, 2]$ with
the 512 Fourier modes in the $x$ direction and time-step $\Delta t = 0.004$. The training data-set used in  the PINN consists of randomly chosen $N_{int}=70$ points from the initial data $u(x,0)$, $N_{end}=0$ point from the end data $u(x,2)$,  $N_b=10$ points pertaining to the periodic boundary data, and $N_f=2000$ points from the spatio-temporal solution zone. The hidden solution $\widehat u(x,t)$ is learned by a 6-hidden-layer deep PINN $f_{nov}(x,t)$ with 20 neurons pere layer combined with a hyperbolic tangent activation function.

Figures~\ref{fnov-bp}(a2) and (c2) exhibit the 2D and 3D profiles of the latent solution $\widehat u (x,t)$ learned by a 6-hidden-layer deep PINN $f_{nov}(x,t)$ with 20 neurons per layer combined with a hyperbolic tangent activation function. The comparison of learning solution (red dashed line), numerical solution (blue solid line) and exact solution (green plus) is shown at three different times $t = 0.5, 0.8$ and $1.6$ (see Fig.~\ref{fnov-bp}(b2)). The $\mathbb{L}_{2}$-norm error between learning solution $\widehat{u}(x,t)$  and numerical solution is 4.56e-02.

\subsection{The generalized $b$-family equation with quartic nonlinearity}

In the subsection, we use the PINN to consider the data-driven peakon solutions of the initial-boundary value problem of the generalized $b$-family equation (\ref{gbf}) with quartic nonlinearity for $b=k=3$.

\begin{figure}[!t]
\begin{center}
\vspace{-0.15in}
\hspace{0.2in} {\scalebox{0.75}[0.75]{\includegraphics{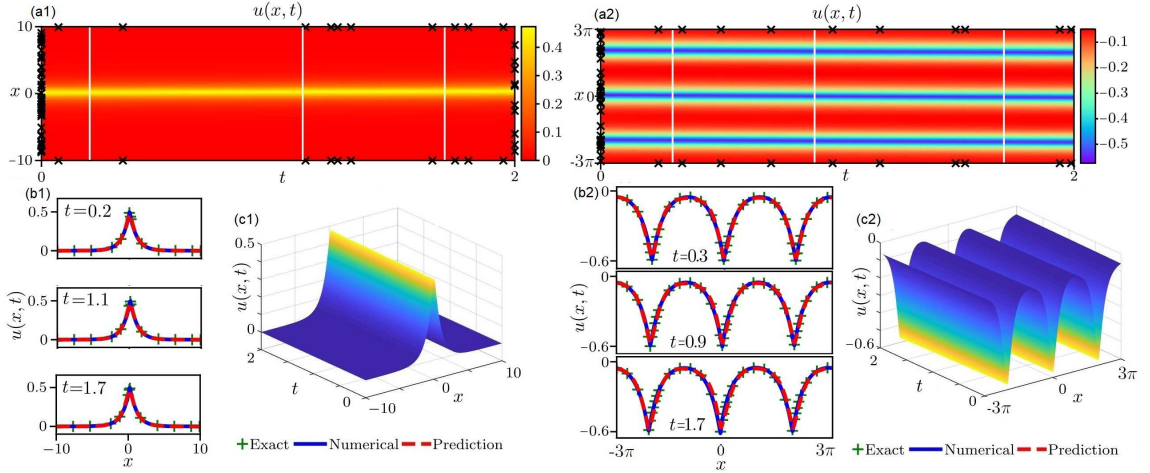}}}
\end{center}
\par
\vspace{-0.05in}
\caption{{\protect\small The generalized b-family equation (\ref{gbf}) with $b=k=3$. The data-driven peakon (a1,c1) and periodic peakon (a2, c2) solutions resulted from the PINN; (b1, b2) The comparisons between the learning, numerical, and exact peakon or periodic peakon solutions at the distinct times (b1) $t=0.2,\, 1.1$, and $1.7$, or (b2) $t=0.3,\, 0.9$, and $1.7$. The $\mathbb{L}_{2}$-norm errors  between the learning and numerical solutions are (b2) 4.62e-02 and (b2) 6.2e-02. $c=0.125$ for (a1,b1,c1), and $c=-0.216$ for (a2,b2,c2).}}
\label{fgbf-bp}
\end{figure}

\v {\it Case 1.} The generalized $b$-family equation (\ref{gbf}) with $b=k=3$ has the peakon solution~\cite{gbf2} $u(x,t)=c^{1/3}e^{-|x-ct|}$. When $c>0$, the peakon solution is a {\it right-going} travelling wave solution, and bright peakon solution, whereas $c<0$, it is a  {\it left-going} travelling wave solution, and dark peakon solution. We here consider  the stationary peakon solution at $t=0$ as the initial condition
\bee \label{gbf-int}
u(x,0)=c^{1/3}\,e^{-|x|},
\ene
and the periodic boundary $u(-L, t)=u(L, t)$. The considered PINN $f_{gbf}(x,t)$ is written as
\bee
f_{gbf}(x,t):=u_t-u_{xxt}+4u^3u_x-3u^2u_xu_{xx}-u^3u_{xxx}.
\ene

We here consider $c=0.125$ in the initial condition (\ref{gbf-int}), and use the Fourier pseudo-spectral method (i.e., one can take Fourier transform in space, and choose the explicit fourth-order Runge-Kutta method in time) to simulate the  generalized $b$-family equation  (\ref{gbf}) with $b=k=3$, the initial value condition (\ref{gbf-int}), and periodic boundary condition. The spatial region is $x\in [-10,\, 10]$ with 256 Fourier modes, and the temporal region is $t\in [0, 2]$ with time-step $\Delta t = 0.004$. As a result, we produce the data-set for the PINN pertaining to the generalized $b$-family equation (\ref{gbf}) with quartic nonlinearity for $b=k=3$. The training data-set used in  the PINN $f_{gbf}(x,t)$ consists of randomly chosen $N_{int}=40$ points from the initial data $u(x,0)$,
$N_{end}=10$ points from the end data $u(x,2)$,  $N_b=10$ points pertaining to the periodic boundary data, and $N_f=3000$ points
from the spatio-temporal solution zone. Figures~\ref{fgbf-bp}(a1) and (c1) exhibit the 2D and 3D profiles of the latent solution $\widehat u (x,t)$ learned by a 5-hidden-layer deep PINN $f_{gbf}(x,t)$ with 20 neurons per layer combined with a hyperbolic tangent activation function. The comparison of learning solution (red dashed line), numerical solution (blue solid line) and exact solution (green plus) is shown at three different times $t = 0.2, 1.1$ and $1.7$ (see Fig.~\ref{fgbf-bp}(b1)). The $\mathbb{L}_{2}$-norm error between learning solution $\widehat{u}(x,t)$  and numerical solution is 4.62e-02.

\v {\it Case 2.} The generalized $b$-family equation (\ref{gbf}) with $b=k=3$ has the periodic peakon solution~\cite{gbf2} $u(x,t)=c^{1/3}\,{\rm sech} \pi \cosh(x-ct-2\pi\lfloor(x-ct)/(2\pi)\rfloor-\pi)$, we consider the stationary periodic peakon solution at $t=0$ as the initial value condition
\bee\label{nov-p}
u(x,0)=c^{1/3}\,{\rm sech}\, \pi \cosh\left(x-2\pi\left\lfloor\frac{x}{2\pi}\right\rfloor-\pi\right),
\ene
and the periodic boundary $u(L_1, t)=u(L_2, t)$, as well as $c=-0.216$. Similarly, we use the pseudo-spectral approach to generate
the data-set for the PINN pertaining to the generalized $b$-family equation (\ref{gbf}) with $b=k=3$ in the spatio-temporal region $(x,t)\in [-3\pi, 3\pi]\times [0, 2]$ with
the 512 Fourier modes in the $x$ direction and time-step $\Delta t = 0.004$. The training data-set used in  the PINN consists of randomly chosen $N_{int}=50$ points from the initial data $u(x,0)$, $N_{end}=0$ point from the end data $u(x,2)$,  $N_b=10$ points pertaining to the periodic boundary data, and $N_f=2000$ points from the spatio-temporal solution zone. The hidden solution $\widehat u(x,t)$ is learned by a 6-hidden-layer deep PINN $f_{gbf}(x,t)$ with 20 neurons per layer combined with a hyperbolic tangent activation function.

Figures~\ref{fgbf-bp}(a2) and (c2) exhibit the 2D and 3D profiles of the latent solution $\widehat u (x,t)$ learned by a 6-hidden-layer deep PINN $f_{gbf}(x,t)$ with 20 neurons per layer combined with a hyperbolic tangent activation function. The comparison of learning solution (red dashed line), numerical solution (blue solid line) and exact solution (green plus) is shown at three different times $t = 0.3, 0.9$ and $1.7$ (see Fig.~\ref{fgbf-bp}(b2)). The $\mathbb{L}_{2}$-norm error between learning solution $\widehat{u}(x,t)$  and numerical solution is 6.2e-02.

\subsection{The mCH equation with cubic nonlinearity}

In the subsection, we use the PINN to consider the data-driven peakon solutions of the initial-boundary value problem of the mCH equation (\ref{mch}).

\begin{figure}[t]
\begin{center}
\vspace{-0.15in}
\hspace{0.2in} {\scalebox{0.75}[0.75]{\includegraphics{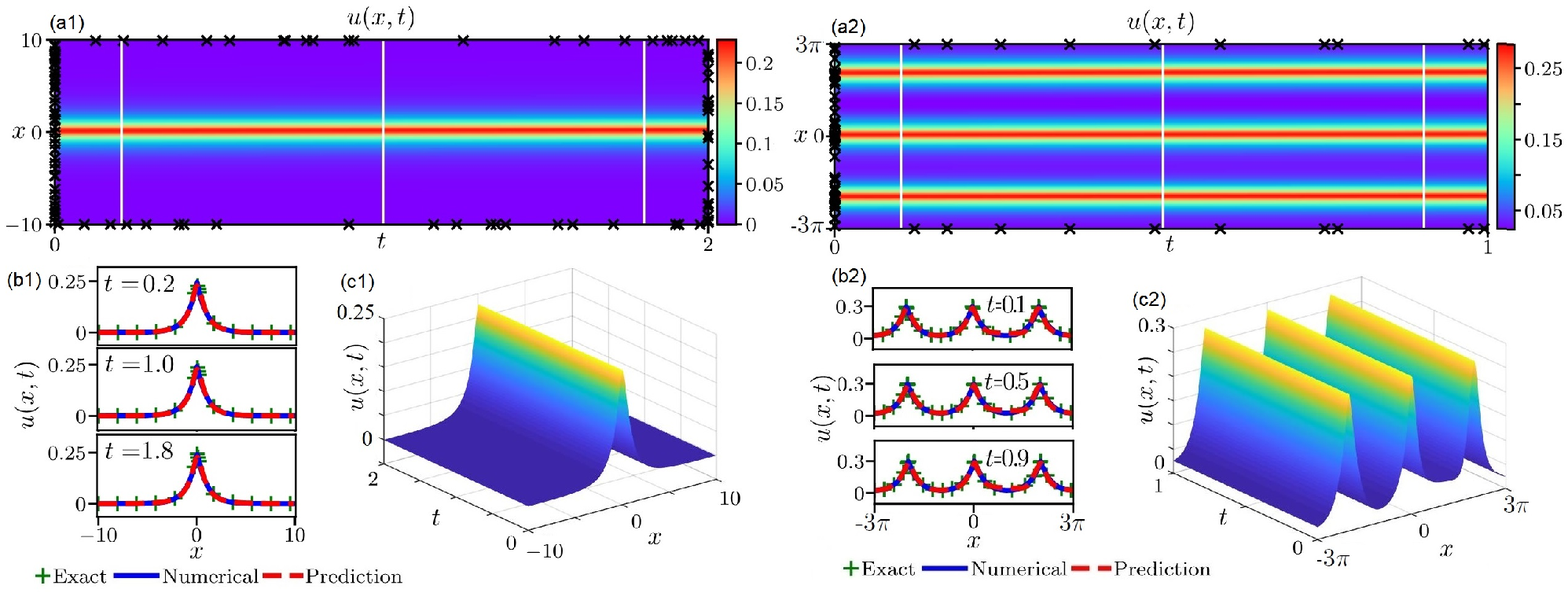}}}
\end{center}
\par
\caption{{\protect\small  The mCH equation (\ref{mch}).  The data-driven peakon (a1,c1) and periodic peakon (a2, c2) solutions resulted from the PINN; (b1, b2) The comparisons between the learning, numerical, and exact peakon or periodic peakon solutions at the distinct times (b1) $t=0.2,\, 1$, and $1.8$, or (b2) $t=0.1,\, 0.5$, and $0.9$. The $\mathbb{L}_{2}$-norm errors  between the learning and numerical solutions are (b2) 4.74e-02 and (b2) 4.09e-02. $c=0.04$ for (a1,b1,c1), and $c=0.06$ for (a2,b2,c2).}}
\label{fmch-bp}
\end{figure}

\v {\it Case 1.} For the peakon solution~\cite{mch-p} $u(x,t)=\sqrt{3c/2}\,e^{-|x-ct|}$ with $c>0$ of the mCH equation (\ref{mch}), which is a {\it right-going} travelling wave solution, and bright peakon solution. We here consider the the stationary peakon solution at $t=0$ as the initial condition
\bee \label{mch-int}
u(x,0)=\sqrt{3c/2}\,e^{-|x|},\quad x\in [-L, L],
\ene
and the periodic boundary $u(-L, t)=u(L, t)$. The considered PINN $f_{mch}(x,t)$ is written as
\bee
 f_{mch}(x,t):=u_t-u_{xxt}+[(u^2-u_x^2)(u-u_{xx})]_x.
\ene

We here consider $c=0.04$ in the initial condition (\ref{mch-int}), and use the Fourier pseudo-spectral method (i.e., one can take Fourier transform in space, and choose the explicit fourth-order Runge-Kutta method in time) to simulate the mCH equation (\ref{mch}) with the initial value condition (\ref{mch-int}) and periodic boundary condition. The spatial region is $x\in [-10,\, 10]$ with 256 Fourier modes, and the temporal region is $t\in [0, 2]$ with time-step $\Delta t = 0.005$. As a result, we produce the data-set for the PINN pertaining to the mCH equation. The training data-set used in  the PINN $f_{mch}(x,t)$ consists of randomly chosen $N_{int}=40$ points from the initial data $u(x,0)$,
$N_{end}=20$ points from the end data $u(x,2)$,  $N_b=20$ points pertaining to the periodic boundary data, and $N_f=2000$ points
from the spatio-temporal solution zone. Figures~\ref{fmch-bp}(a1) and (c1) exhibit the 2D and 3D profiles of the latent solution $\widehat u (x,t)$ learned by a 7-hidden-layer deep PINN $f_{mch}(x,t)$ with 20 neurons per layer combined with a hyperbolic tangent activation function. The comparison of learning solution (red dashed line), numerical solution (blue solid line) and exact solution (green plus) is shown at three different times $t = 0.2, 1$, and $1.8$ (see Fig.~\ref{fmch-bp}(b1)).  The $\mathbb{L}_{2}$-norm error between learning solution $\widehat{u}(x,t)$  and numerical solution is 4.74e-02.

\v {\it Case 2.} The mCH equation has the periodic peakon solution~\cite{mch-pp} $ u(x,t)=\sqrt{3c/(2\cosh^2\pi+1)}\cosh(x-ct-2\pi\lfloor(x-ct)/(2\pi)\rfloor-\pi$ with $ c>0$. We consider the stationary periodic peakon solution at $t=0$ as the initial value condition
\bee
 u(x,0)=\sqrt{\frac{3c}{2\cosh^2\pi+1}}\cosh\left(x-2\pi\left\lfloor\frac{x}{2\pi}\right\rfloor-\pi\right), \quad c>0
\ene
and the periodic boundary $u(L_1, t)=u(L_2, t)$, as well as $c=0.06$. Similarly, we use the pseudo-spectral approach to generate
the data-set for the PINN pertaining to the mCH equation (\ref{mch}) in the spatio-temporal region $(x,t)\in [-3\pi, 3\pi]\times [0, 1]$ with
the 512 Fourier modes in the $x$ direction and time-step $\Delta t = 0.002$. The training data-set used in  the PINN consists of randomly chosen $N_{int}=60$ points from the initial data $u(x,0)$, $N_{end}=0$ point from the end data $u(x,1)$,  $N_b=10$ points pertaining to the periodic boundary data, and $N_f=2000$ points from the spatio-temporal solution zone. The hidden solution $\widehat u(x,t)$ is learned by a 6-hidden-layer deep PINN $f_{mch}(x,t)$ with 20 neurons per layer combined with a hyperbolic tangent activation function.

Figures~\ref{fmch-bp}(a2) and (c2) exhibit the 2D and 3D profiles of the latent solution $\widehat u (x,t)$ learned by a 6-hidden-layer deep PINN $f_{mch}(x,t)$ with 20 neurons per layer combined with a hyperbolic tangent activation function. The comparison of learning solution (red dashed line), numerical solution (blue solid line) and exact solution (green plus) is shown at three different times $t = 0.1, 0.5$ and $0.9$ (see Fig.~\ref{fmch-bp}(b2)). The $\mathbb{L}_{2}$-norm error between learning solution $\widehat{u}(x,t)$  and numerical solution is 4.09e-02.

\subsection{The mCH-CH equation with both quadratic and cubic nonlinearities}

{\it Case 1.} The mCH-CH equation (\ref{mch2}) has the peakon solution~\cite{mch-ch} $u(x,t)=(-3k_2\pm \sqrt{9k_2^2+24ck_1})/(4k_1)e^{-|x-ct|}$,
which is a {\it right-going} ($c>0$ and $9k_2^2+24ck_1>0$) or {\it left-going} ($c<0$ and $9k_2^2+24ck_1>0$) peaked travelling wave solution. We here consider the the stationary peakon solution at $t=0$ as the initial condition
\bee \label{mch2-int}
u(x,0)=\frac{-3k_2+ \sqrt{9k_2^2+24ck_1}}{4k_1}e^{-|x|},
\ene
and the periodic boundary $u(-L, t)=u(L, t)$. The considered PINN $f_{mch2}(x,t)$ is written as
\bee
 f_{mch2}(x,t):=u_t-u_{xxt}+k_1[(u^2-u_x^2)(u-u_{xx})]_x+k_2(3uu_x-2u_xu_{xx}-uu_{xxx}).
\ene

i) For $c=0.5,\, k_1=-0.2$ and $k_2=2$ in the initial condition (\ref{mch2-int}), which is a bright peakon solution, we use the Fourier pseudo-spectral method (i.e., one can take Fourier transform in space, and choose the explicit fourth-order Runge-Kutta method in time) to simulate the mCH-CH equation (\ref{mch2}) with the initial value condition (\ref{mch2-int}) and periodic boundary condition. The spatial region is $x\in [-10,\, 10]$ with 640 Fourier modes, and the temporal region is $t\in [0, 2]$ with time-step $\Delta t = 0.004$. As a result, we produce the data-set for the PINN pertaining to the mCH-CH equation. The training data-set used in  the PINN $f_{mch2}(x,t)$ consists of randomly chosen $N_{int}=40$ points from the initial data $u(x,0)$, $N_{end}=10$ points from the end data $u(x,2)$,  $N_b=10$ points pertaining to the periodic boundary data, and $N_f=5000$ points from the spatio-temporal solution zone. Figures~\ref{fmch2-bd}(a1) and (c1) exhibit the 2D and 3D profiles of the latent solution $\widehat u (x,t)$ learned by a 5-hidden-layer deep PINN $f_{mch2}(x,t)$ with 10 neurons per layer combined with a hyperbolic tangent activation function. The comparison of learning solution (red dashed line), numerical solution (with blue solid line) and exact solution (green plus) is shown at three different times $t = 0.2, 0.7$, and $1.4$ (see Fig.~\ref{fmch2-bd}(b1)). The $\mathbb{L}_{2}$-norm error between learning solution $\widehat{u}(x,t)$  and numerical solution is 6.02e-02.

ii) For $c=-2,\, k_1=-0.6$ and $k_2=4$ in the initial condition (\ref{mch2-int}), which is a dark peakon solution, we similarly choose the spatial region $x\in [-10,\, 10]$ with 640 Fourier modes, and the temporal region is $t\in [0, 2]$ with time-step $\Delta t = 0.004$.  As a result, we produce the data-set for the PINN pertaining to the mCH-CH equation. The training data-set used in  the PINN $f_{mch2}(x,t)$ consists of randomly chosen $N_{int}=40$ points from the initial data $u(x,0)$, $N_{end}=10$ points from the end data $u(x,2)$,  $N_b=10$ points pertaining to the periodic boundary data, and $N_f=6000$ points from the spatio-temporal solution zone. Figures~\ref{fmch2-bd}(a2) and (c2) exhibit the 2D and 3D profiles of the latent solution $\widehat u (x,t)$ learned by a 6-hidden-layer deep PINN $f_{mch2}(x,t)$ with 10 neurons per layer combined with a hyperbolic tangent activation function. The comparison of learning solution (red dashed line), numerical solution (blue solid line) and exact solution (green plus) is shown at three different times $t = 0.1, 0.6$, and $1.3$ (see Fig.~\ref{fmch2-bd}(b2)). The $\mathbb{L}_{2}$-norm error between learning solution $\widehat{u}(x,t)$  and numerical solution is 8.47e-02.
\begin{figure}[!t]
\begin{center}
\vspace{-0.1in}
\hspace{0.2in} {\scalebox{0.75}[0.85]{\includegraphics{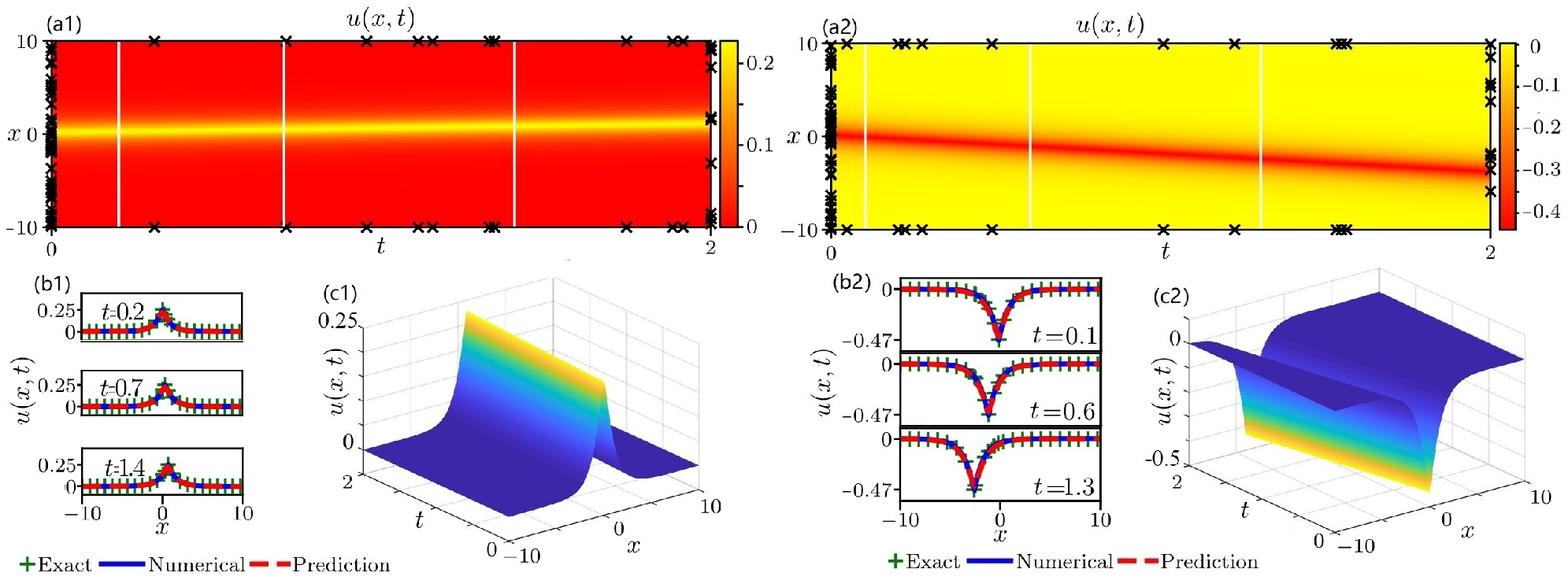}}}
\end{center}
\par
\caption{{\protect\small The mCH-CH equation with peakon solutions: (a1,a2) The data-driven peakon solution resulted from the PINN; (b1,b2) The comparisons between the learning, numerical, and exact peakon solutions at the distinct times (b1) $t = 0.2,\, 0.7$, and $1.4$, (b2)
$t = 0.1,\, 0.6$, and $1.3$. The $\mathbb{L}_{2}$-norm errors $\widehat u(x,t)$ between learning and numerical peakon solutions are (b1) 6.02e-02 and (b2) 8.47e-02; (c1,c2) The 3D profiles of the learning bright and dark peakon solutions. $c=0.5,\, k_1=-0.2,\, k_2=2$ for (a1, b1, c1) and $c=-2,\, k_1=-0.6,\, k_2=4$ for (a2, b2, c2).}}
\label{fmch2-bd}
\end{figure}

\v {\it Case 2.} The mCH-CH equation has the periodic peakon solution $ u(x,t)=(-3k_2\cosh\pi\pm a)/[2k_1(2\cosh^2\pi+1)]
 \cosh(x-ct-2\pi\lfloor(x-ct)/(2\pi)\rfloor-\pi)$ with $a=\sqrt{9k_2^2\cosh^2\pi+12ck_1(2\cosh^2\pi+1)}$.
We here consider the the stationary periodic peakon solution at $t=0$ as the initial condition
\bee\label{mch2-p}
 u(x,0)=\frac{-3k_2\cosh\pi\!+\!\mu\sqrt{9k_2^2\cosh^2\!\pi\!+\!12ck_1(2\cosh^2\!\pi\!+\!1)}}{2k_1(2\cosh^2\!\pi+1)}
 \cosh\left(x\!-\!2\pi\left\lfloor\frac{x}{2\pi}\right\rfloor\!-\!\pi\right),\,\,\,\, \mu=\pm 1
\ene
and the periodic boundary condition $u(L_1, t)=u(L_2, t)$

i) We consider $\mu=1,\, c=1.5,\,k_1=-0.5$, and $k_2=2$ in the initial condition (\ref{mch2-p}), and use the pseudo-spectral approach to generate the data-set for the PINN pertaining to the mCH-CH equation in the spatio-temporal region $(x,t)\in [-75\pi/32, 117\pi/32]\times [0, 2]$ with
the 640 Fourier modes in the $x$ direction and time-step $\Delta t = 0.004$. The training data-set used in  the PINN consists of randomly chosen $N_{int}=90$ points from the initial data $u(x,0)$, $N_{end}=0$ point from the end data $u(x,2)$,  $N_b=10$ points pertaining to the periodic boundary data, and $N_f=6000$ points from the spatio-temporal solution zone. The hidden solution $\widehat u(x,t)$ is learned by a 6-hidden-layer deep PINN $f_{mch2}(x,t)$ with 40 neurons per layer combined with a hyperbolic tangent activation function.

Figures~\ref{fmch2-p}(a1) and (c1) exhibit the 2D and 3D profiles of the latent solution $\widehat u (x,t)$ learned by a 6-hidden-layer deep PINN $f_{mch2}(x,t)$ with 40 neurons per layer combined with a hyperbolic tangent activation function. The comparison of learning solution (red dashed line), numerical solution (blue solid line) and exact solution (green plus) is shown at three different times $t = 0.3, 0.8$ and $1.7$ (see Fig.~\ref{fmch2-p}(b1)).  The $\mathbb{L}_{2}$-norm error between learning solution $\widehat{u}(x,t)$  and numerical solution is 7.35e-02.

\begin{figure}[!t]
\begin{center}
\vspace{-0.15in}
\hspace{0.2in} {\scalebox{0.75}[0.75]{\includegraphics{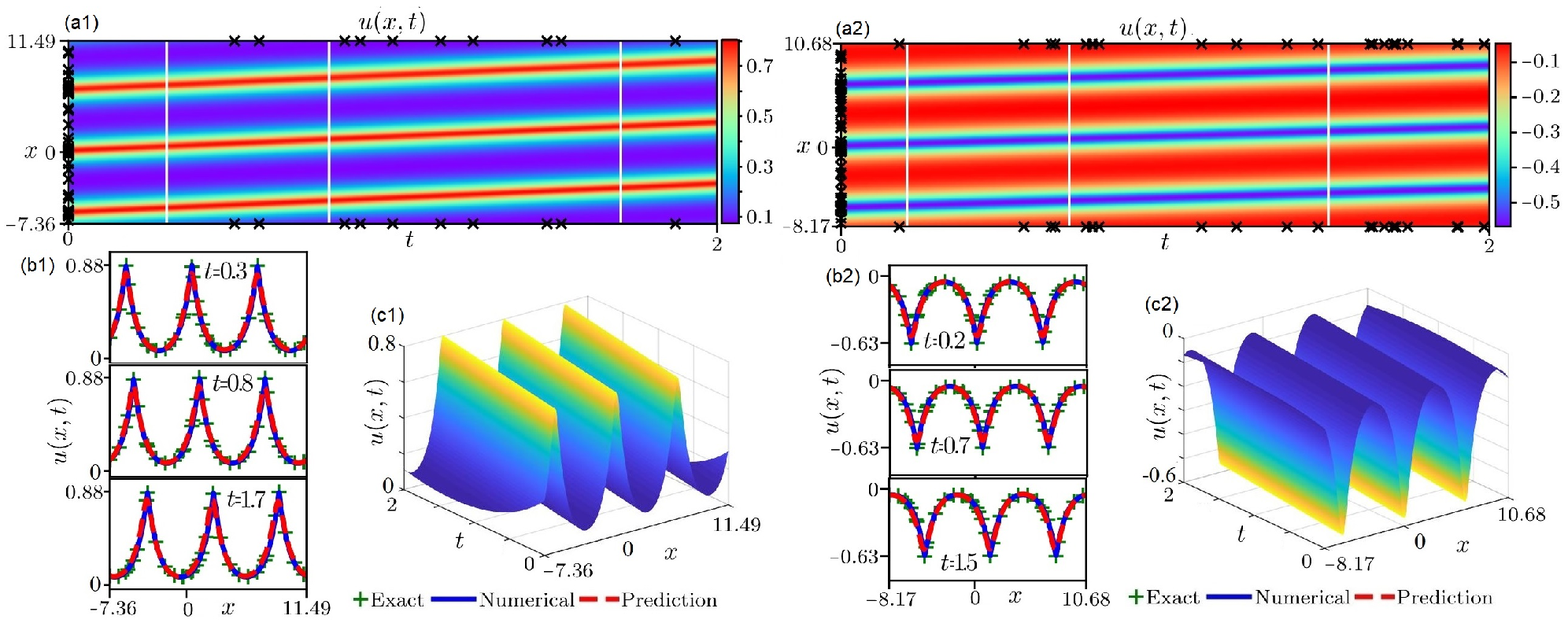}}}
\end{center}
\par
\caption{{\protect\small The mCH-CH equation. (a1,a2) The data-driven periodic peakon solution resulted from the PINN; (b1,b2) The comparisons between the learning, numerical, and exact periodic  peakon solutions at the distinct times (b1) $t = 0.3,\, 0.8$, and $1.7$, (b2) $t = 0.2,\, 0.7$, and $1.5$. The $\mathbb{L}_{2}$-norm errors $\widehat u(x,t)$ between learning and numerical peakon solutions are (b1) 7.35e-02 and (b2) 7.99e-02; (c1,c2) The 3D profiles of the learning bright and dark peakon solutions. $\mu=1,\, c=1.5,\,k_1=-0.5,\, k_2=2$ for (a1, b1, c1) and $\mu=-1,\, c=1,\,k_1=-1,\, k_2=-2$ for (a2, b2, c2).}}
\label{fmch2-p}
\end{figure}

ii) Similarly, we consider $\mu=-1,\, c=1,\,k_1=-1$, and $k_2=-2$ in the initial condition (\ref{mch2-p}), use the pseudo-spectral approach to generate
the data-set for the PINN pertaining to the mCH-CH equation in the spatio-temporal region $(x,t)\in [-2.6\pi, 3.4\pi]\times [0, 2]$ with
the 640 Fourier modes in the $x$ direction and time-step $\Delta t = 0.004$. The training data-set used in  the PINN consists of randomly chosen $N_{int}=90$ points from the initial data $u(x,0)$, $N_{end}=0$ points from the end data $u(x,2)$,  $N_b=20$ points pertaining to the periodic boundary data, and $N_f=6000$ points from the spatio-temporal solution zone. The hidden solution $\widehat u(x,t)$ is learned by a 6-hidden-layer deep PINN $f_{mch2}(x,t)$ with 40 neurons per layer combined with a hyperbolic tangent activation function.

Figures~\ref{fmch2-p}(a2) and (c2) exhibit the 2D and 3D profiles of the latent solution $\widehat u (x,t)$ learned by a 6-hidden-layer deep PINN $f_{mch2}(x,t)$ with 40 neurons per layer combined with a hyperbolic tangent activation function. The comparison of learning solution (red dashed line), numerical solution (blue solid line) and exact solution (green plus) is shown at three different times $t = 0.2,\, 0.7$ and $1.5$ (see Fig.~\ref{fmch2-p}(b2)).  The $\mathbb{L}_{2}$-norm error between learning solution $\widehat{u}(x,t)$  and numerical solution is 7.99e-02.

\begin{figure}[!t]
\begin{center}
\vspace{-0.1in}
\hspace{0.2in} {\scalebox{0.72}[0.72]{\includegraphics{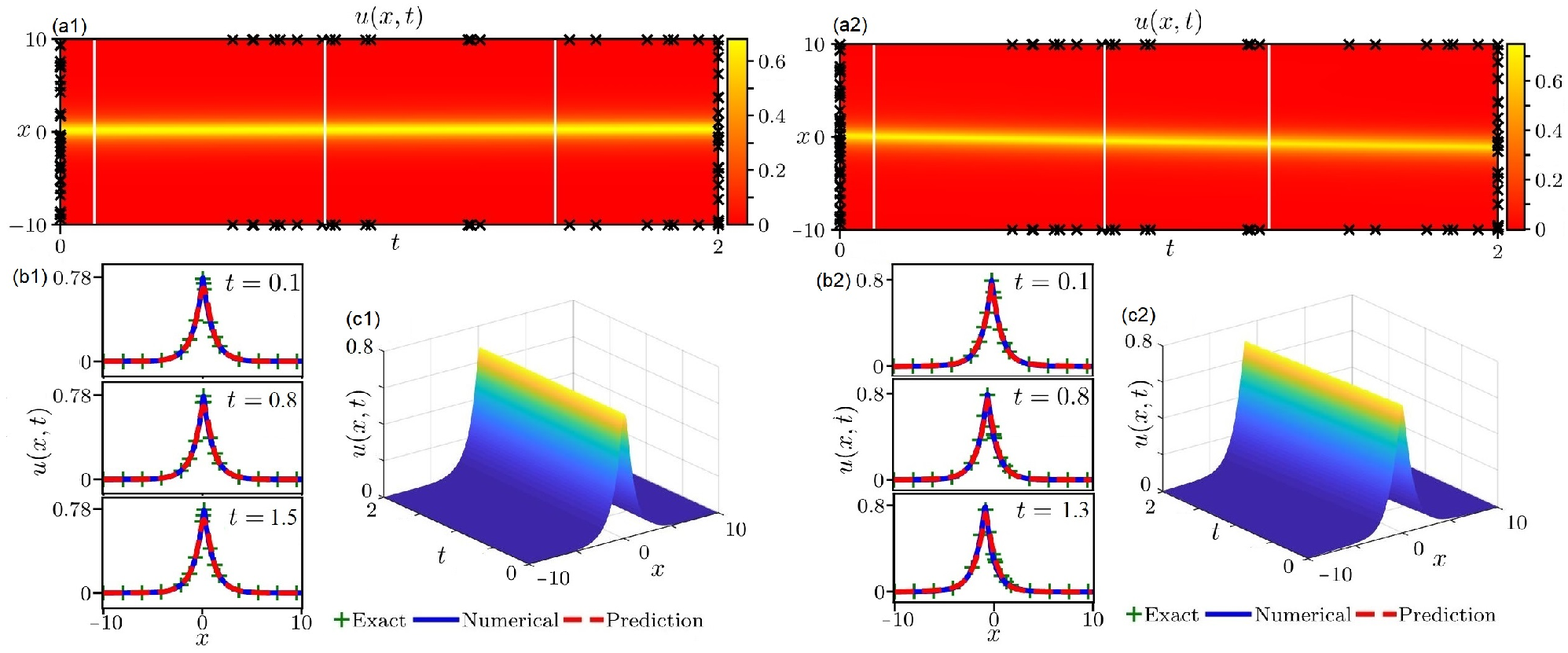}}}
\end{center}
\par
\caption{{\protect\small The mCH-Novikov equation. (a1,a2) The data-driven peakon solution resulted from the PINN; (b1,b2) The comparisons between the learning, numerical, and exact peakon solutions at the distinct times (b1) $t = 0.1,\, 0.8$, and $1.5$, (b2) $t = 0.1,\, 0.8$, and $1.3$. The $\mathbb{L}_{2}$-norm errors $\widehat u(x,t)$ between learning and numerical peakon solutions are (b1) 6.02e-02 and (b2) 1.31e-02; (c1,c2) The 3D profiles of the learning bright peakon solutions. $c=0.1,\, k_1=-0.25,\, k_2=1/3$ for (a1, b1, c1) and $c=-0.7,\, k_1=-0.15,\, k_2=-1$ for (a2, b2, c2).}}
\label{fmchn-bd}
\end{figure}

\subsection{The mCH-Novikov equation with cubic nonlinearity}

In the subsection, we use the PINN to consider the data-driven peakon and periodic peakon solutions of the initial-boundary value problem of the
mCH-Novikov equation (\ref{mchn}).

\v {\it Case 1.} The mCH-Novikov equation (\ref{mchn}) has the peakon solution~\cite{mch-n} $u(x,t)=\sqrt{3c/(2k_1+3k_2)}e^{-|x-ct|}$, which is
a {\it right-going} for $c>0,\,2k_1+3k_2>0$ or {\it left-going} for $c<0,\,2k_1+3k_2<0$ peaked travelling wave solution. We here consider the the stationary peakon solution at $t=0$ as the initial condition
\bee \label{mchn-int}
u(x,0)=\sqrt{3c/(2k_1+3k_2)}e^{-|x|},\quad c>0,
\ene
and the periodic boundary $u(-L, t)=u(L, t)$. The considered PINN $f_{mchn}(x,t)$ is written as
\bee
 f_{mchn}(x,t):=u_t-u_{xxt}+k_1[(u^2-u_x^2)(u-u_{xx})]_x+k_2(4u^2u_x-3uu_xu_{xx}-u^2u_{xxx}).
\ene

i) For $c=0.1,\, k_1=-0.25$ and $k_2=1/3$ in the initial condition (\ref{mchn-int}), which is a bright peakon solution, we use the Fourier pseudo-spectral method (i.e., one can take Fourier transform in space, and choose the explicit fourth-order Runge-Kutta method in time) to simulate the mCH-Novikov equation (\ref{mchn}) with the initial value condition (\ref{mchn-int}) and periodic boundary condition. The spatial region is $x\in [-10,\, 10]$ with 256 Fourier modes, and the temporal region is $t\in [0, 2]$ with time-step $\Delta t = 0.004$. As a result, we produce the data-set for the PINN pertaining to the mCH-Novikov equation. The training data-set used in  the PINN $f_{mchn}(x,t)$ consists of randomly chosen $N_{int}=30$ points from the initial data $u(x,0)$, $N_{end}=20$ points from the end data $u(x,2)$,  $N_b=20$ points pertaining to the periodic boundary data, and $N_f=1000$ points from the spatio-temporal solution zone. Figures~\ref{fmchn-bd}(a1) and (c1) exhibit the 2D and 3D profiles of the latent solution $\widehat u (x,t)$ learned by a 7-hidden-layer deep PINN $f_{mchn}(x,t)$ with 10 neurons per layer combined with a hyperbolic tangent activation function. The comparison of learning solution (red dashed line), numerical solution (blue solid line) and exact solution (green plus) is shown at three different times $t = 0.1,\, 0.8$, and $1.5$ (see Fig.~\ref{fmchn-bd}(b1)). The $\mathbb{L}_{2}$-norm error between learning solution $\widehat{u}(x,t)$  and numerical solution is 6.02e-02.

ii) We here consider $c=-0.7,\, k_1=-0.15$ and $k_2=-1$ in the initial condition (\ref{mchn-int}) such that the initial condition is a bright peakon. Similarly, we choose the spatial region $x\in [-10,\, 10]$ with 256 Fourier modes, and the temporal region is $t\in [0, 2]$ with time-step $\Delta t = 0.004$.  As a result, we produce the data-set for the PINN pertaining to the mCH-Novikov equation. The training data-set used in  the PINN $f_{mchn}(x,t)$ consists of randomly chosen $N_{int}=50$ points from the initial data $u(x,0)$, $N_{end}=20$ points from the end data $u(x,2)$,  $N_b=20$ points pertaining to the periodic boundary data, and $N_f=1000$ points from the spatio-temporal solution zone. Figures~\ref{fmchn-bd}(a2) and (c2) exhibit the 2D and 3D profiles of the latent solution $\widehat u (x,t)$ learned by a 6-hidden-layer deep PINN $f_{mchn}(x,t)$ with 10 neurons per layer combined with a hyperbolic tangent activation function. The comparison of learning solution (red dashed line), numerical solution (with blue solid line) and exact solution (green plus) is shown at three different times $t = 0.1,\, 0.8$, and $1.3$ (see Fig.~\ref{fmchn-bd}(b2)). The $\mathbb{L}_{2}$-norm error between learning solution $\widehat{u}(x,t)$  and numerical solution is 1.31e-02.

\begin{figure}[!t]
\begin{center}
\vspace{-0.15in} {\scalebox{0.45}[0.35]{\includegraphics{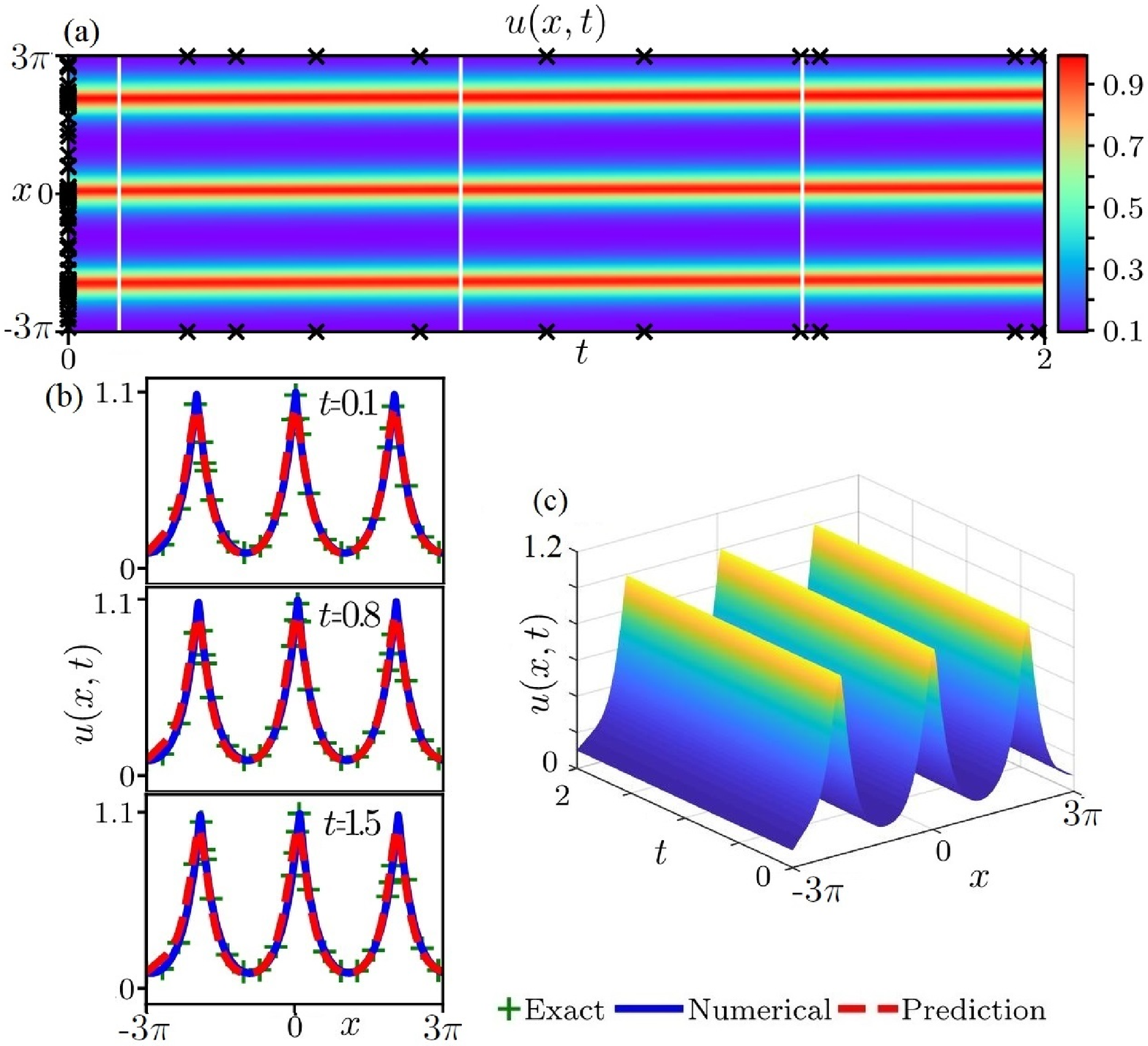}}}
\end{center}
\vspace{-0.15in}
\caption{{\protect\small  The mCH-Novikov equation. (a) The data-driven periodic peakon solution resulted from the PINN; (b) The comparisons between the learning, numerical, and exact periodic peakon solutions at the distinct times $t = 0.1$, $0.8$, and $1.5$. The $\mathbb{L}_{2}$-norm error  between the learning and numerical periodic peakon solutions is  8.23e-02; (c) The 3D profile of the learning periodic peakon solution.}}
\label{fmchn-p}
\end{figure}

\v {\it Case 2.} The mCH-Novikov equation (\ref{mchn}) can be shown to possess the periodic peakon solution $ u(x,t)=\sqrt{3c/[k_1(2\cosh^2\pi+1)+3k_2\cosh^2\pi]}\cosh(x-ct-2\pi\lfloor(x-ct)/(2\pi)\rfloor-\pi$ with $ c>0$.
We consider the stationary periodic peakon solution at $t=0$ as the initial value condition
\bee \label{mchn-p}
 u(x,0)=\sqrt{\frac{3c}{k_1(2\cosh^2\pi+1)+3k_2\cosh^2\pi}}\cosh\left(x-2\pi\left\lfloor\frac{x}{2\pi}\right\rfloor-\pi\right), \quad c>0
\ene
and the periodic boundary condition $u(L_1, t)=u(L_2, t)$, as well as $c=0.2,\,k_1=-0.25,\, k_2=1/3$.

Similarly, we use the pseudo-spectral approach to generate
the data-set for the PINN pertaining to the mCH-Novikov equation in the spatio-temporal region $(x,t)\in [-3\pi, 3\pi]\times [0, 2]$ with
the 512 Fourier modes in the $x$ direction and time-step $\Delta t = 0.004$. The training data-set used in  the PINN consists of randomly chosen $N_{int}=60$ points from the initial data $u(x,0)$, $N_{end}=0$ point from the end data $u(x,2)$,  $N_b=10$ points pertaining to the periodic boundary data, and $N_f=2000$ points from the spatio-temporal solution zone. The hidden solution $\widehat u(x,t)$ is learned by a 6-hidden-layer deep PINN $f_{mchn}(x,t)$ with 20 neurons per layer combined with a hyperbolic tangent activation function.

Figures~\ref{fmchn-p}(a) and (c) exhibit the 2D and 3D profiles of the latent solution $\widehat u (x,t)$ learned by a 6-hidden-layer deep PINN $f_{mchn}(x,t)$ with 20 neurons per layer combined with a hyperbolic tangent activation function. The comparison of learning solution (red dashed line), numerical solution (blue solid line) and exact solution (green plus) is shown at three different times $t = 0.1,\, 0.8$ and $1.5$ (see Fig.~\ref{fmchn-p}(b)). The $\mathbb{L}_{2}$-norm error between learning solution $\widehat{u}(x,t)$  and numerical solution is 8.23e-02.

\begin{figure}[!t]
\begin{center}
\vspace{-0.1in}
\hspace{0.2in} {\scalebox{0.75}[0.75]{\includegraphics{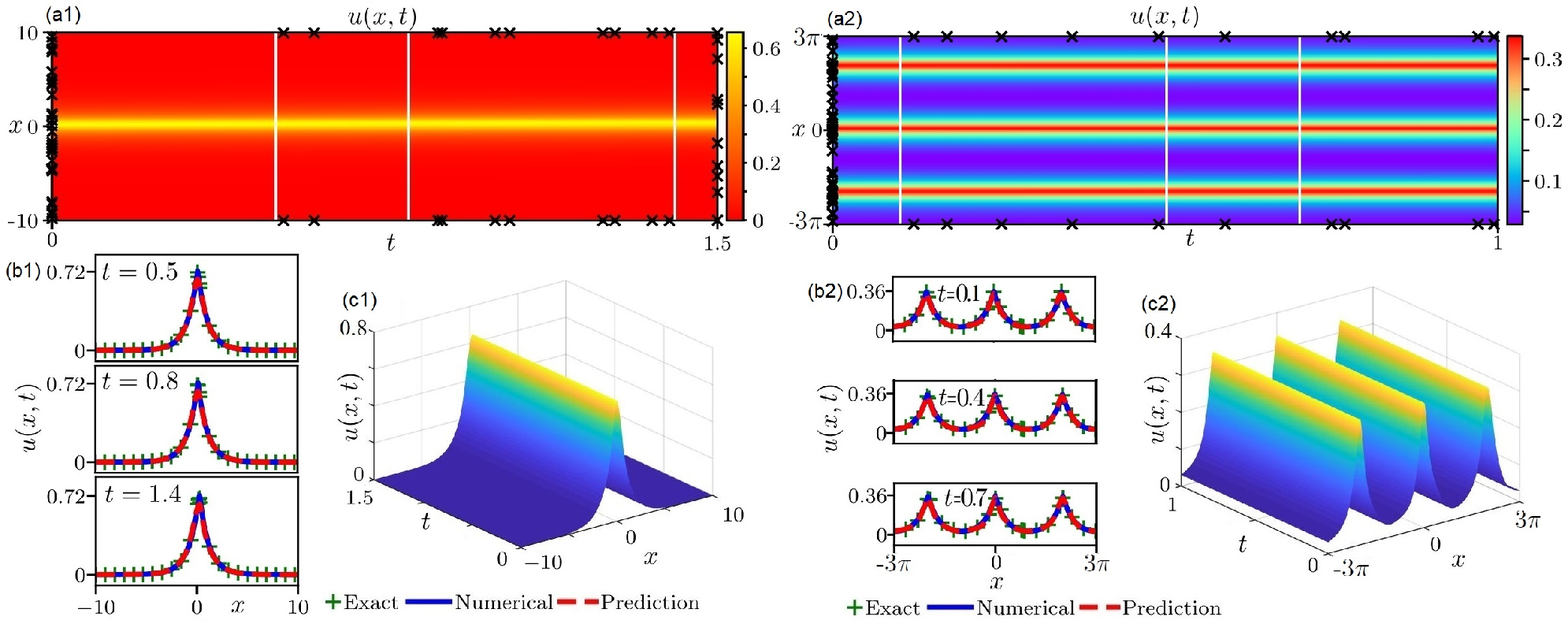}}}
\end{center}
\par
\caption{{\protect\small The generalized mCH equation (\ref{gmch}) with $k=2$: (a1,a2) The data-driven peakon solution resulted from the PINN; (b1, b2) The comparisons between the learning, numerical, and exact peakon solutions at the distinct times (b1) $t=0.5,\, 0.8,\, 1.4$ or (b2) $t=0.1,\, 0.4,\, 0.7$. The $\mathbb{L}_{2}$-norm errors $\widehat u(x,t)$ between learning and numerical peakon solutions are (b1) 5.29e-02 and (b2) 3.84e-02; (c1,c2) The 3D profiles of the learning bright peakon and periodic peakon solutions. $c=0.15$ for (a1, b1, c1) and $c=0.1$ for (a2, b2, c2).}}
\label{fgmch-bp}
\end{figure}

\subsection{The generalized mCH equation with quintic nonlinearity}

In the subsection, we use the PINN to consider the data-driven peakon and periodic peakon solutions of the initial-boundary value problem of
the generalized mCH equation (\ref{gmch}) with $k=2$.

\v {\it Case 1.} The generalized mCH equation (\ref{gmch}) with $k=2$ admits the peakon solution~\cite{gmch-s} $
 u(x,t)=\sqrt[4]{15c/8}\,e^{-|x-ct|}$ with $c>0$. We here consider the the stationary peakon solution at $t=0$ as the initial condition
 \bee\label{gmch-int}
 u(x,0)=\sqrt[4]{\frac{15c}{8}}\,e^{-|x|}, \quad c>0,
 \ene
 and the periodic boundary $u(-L, t)=u(L, t)$. The considered PINN $f_{gmch}(x,t)$ is written as
\bee \label{fgmch}
 f_{gmch}(x,t):=u_t-u_{xxt}+[(u^2-u_x^2)^2(u-u_{xx})]_x.
\ene

We here consider $c=0.15$ in the initial condition (\ref{gmch-int}), and use the Fourier pseudo-spectral method (i.e., one can take Fourier transform in space, and choose the explicit fourth-order Runge-Kutta method in time) to simulate the generalized mCH equation (\ref{gmch}) with $k=2$ and the initial value condition (\ref{gmch-int}) and periodic boundary condition. The spatial region is $x\in [-10,\, 10]$ with 640 Fourier modes, and the temporal region is $t\in [0, 1.5]$ with time-step $\Delta t = 0.003$. As a result, we produce the data-set for the PINN pertaining to the generalized mCH equation. The training data-set used in  the PINN $f_{gmch}(x,t)$ consists of randomly chosen $N_{int}=30$ points from the initial data $u(x,0)$, $N_{end}=10$ points from the end data $u(x, 1.5)$,  $N_b=10$ points pertaining to the periodic boundary data, and $N_f=2000$ points
from the spatio-temporal solution zone. Figures~\ref{fgmch-bp}(a1) and (c1) exhibit the 2D and 3D profiles of the latent solution $\widehat u (x,t)$ learned by a 5-hidden-layer deep PINN $f_{gmch}(x,t)$ with 10 neurons per layer combined with a hyperbolic tangent activation function. The comparison of learning solution (red dashed line), numerical solution (blue solid line) and exact solution (green plus) is shown at three different times $t = 0.5, 0.8$, and $1.4$ (see Fig.~\ref{fgmch-bp}(b1)). The $\mathbb{L}_{2}$-norm error between learning solution $\widehat{u}(x,t)$  and numerical solution is 5.29e-02.

\v {\it Case 2.} The generalized mCH equation (\ref{gmch}) with $k=2$ possesses the periodic peakon solution $ u(x,t)=\sqrt[4]{15c \,{\rm csch}\pi/(8\cosh^4\pi+4\cosh^2\pi+3)}\cosh(x-ct-2\pi\lfloor(x-ct)/(2\pi)\rfloor-\pi)$ with $ c>0$. We consider the stationary periodic peakon solution at $t=0$ as the initial value condition
\bee\label{gmch-int}
 u(x,0)=\sqrt[4]{\frac{15c\,{\rm csch}\pi}{8\cosh^4\pi+4\cosh^2\pi+3}}\cosh\left(x-2\pi\left\lfloor\frac{x}{2\pi}\right\rfloor-\pi\right), \quad c>0
\ene
and the periodic boundary condition $u(-L, t)=u(L, t)$, as well as $c=0.1$.

Similarly, we use the pseudo-spectral approach to generate the data-set for the PINN pertaining to the generalized mCH equation (\ref{gmch}) with $k=2$ in the spatio-temporal region $(x,t)\in [-3\pi, 3\pi]\times [0, 1]$ with
the 512 Fourier modes in the $x$ direction and time-step $\Delta t = 0.002$. The training data-set used in  the PINN consists of randomly chosen $N_{int}=60$ points from the initial data $u(x,0)$, $N_{end}=0$ point from the end data $u(x,2)$,  $N_b=10$ points pertaining to the periodic boundary data, and $N_f=2000$ points from the spatio-temporal solution zone. The hidden solution $\widehat u(x,t)$ is learned by a 6-hidden-layer deep PINN $f_{gmch}(x,t)$ with 20 neurons per layer combined with a hyperbolic tangent activation function.

Figures~\ref{fgmch-bp}(a2) and (c2) exhibit the 2D and 3D profiles of the latent solution $\widehat u (x,t)$ learned by a 6-hidden-layer deep PINN $f_{gmch}(x,t)$ with 20 neurons per layer combined with a hyperbolic tangent activation function. The comparison of learning solution (red dashed line), numerical solution (blue solid line) and exact solution (green plus) is shown at three different times $t = 0.1,\, 0.4$ and $0.7$ (see Fig.~\ref{fgmch-bp}(b2)). The $\mathbb{L}_{2}$-norm error between learning solution $\widehat{u}(x,t)$  and numerical solution is 3.84e-02.

\section{Conclusions and discussions}

In summary, we have applied the PINN deep learning to successfully explore the data-driven peakon and periodic peakon solutions of some nonlinear dispersive equations with some initial-boundary value conditions such as  the Camassa-Holm (CH) equation, Degasperis-Procesi equation, modified CH equation, $b$-family equation, Novikov equation, mCH-Novikov equation, and etc. Moreover, the PINN deep learning approach can also be extended to other nonlinear peakon equations given by Eqs.~(\ref{gpe1}) and (\ref{gpe2}), and higher-dimensional coupled nonlinear dispersive equations with peakon solutions. These questions will be studied in another literature.


\v \noindent {\bf Acknowledgements} \v

This work is partially supported by the NSFC under Grant Nos. 11731014 and 11925108.

\v \noindent {\bf Appendix A.\, The DP equation}

\vspace{0.05in} In the appendix, we use the PINN deep learning approach to consider the data-driven peakon and periodic peakon solutions of the initial-boundary value problem of the DP equation (\ref{dp}). Though the CH equation and DP equation both admit the same peakon and periodic peakon solutions,
but their Lax pairs are different, where the CH equation has the second-order Lax pair~\cite{ch}, and the DP equation admits the third-order Lax pair~\cite{bch1}.

\v {\it Case 1.} Since the DP equation (\ref{dp}) also admit the peakon solution~\cite{bch1} $u(x,t)=ce^{-|x-ct|}$, thus we consider
the initial value condition (\ref{ch-int}) and periodic boundary condition $u(-L, t)=u(L, t)$.  The considered PINN $f_{dp}(x,t)$ is chosen as
\bee \label{fdp}
 f_{dp}(x,t):=u_t-u_{xxt}+4uu_x-3u_xu_{xx}-uu_{xxx}.
\ene

i) We take $c=0.6$ in the initial condition (\ref{ch-int}) and  use the pseudo-spectral method to generate the data-set for the PINN pertaining to the DP equation (\ref{dp}) in the spatio-temporal region $(x,t)\in [-5, 5]\times [0, 3]$ with the 256 Fourier modes in the $x$ direction and time-step $\Delta t = 0.006$. The training data-set used in the deep PINN $f_{dp}(x,t)$ consists of randomly chosen $N_{int}=10$ points from the initial data $u(x,0)$, $N_{end}=10$ points from the end data $u(x,3)$,  $N_b=10$ points pertaining to the periodic boundary data, and $N_f=1000$ points from the spatial-temporal solution zone. The hidden solution $\widehat u(x,t)$ can be trained by the the 5-hidden-layer deep PINN $f_{dp}(x,t)$ with 10 neurons per layer and a hyperbolic tangent activation function.

 Figures~\ref{dp-bd}(a1) and (c1) exhibit the 2D and 3D profiles of the latent bright-peakon solution $\widehat u (x,t)$.
 The comparison of learning solution (red dashed line), numerical solution (blue solid line) and exact solution (green plus)
 is shown at three different times $t = 0.3,\, 1.5$, and $2$ (see Fig.~\ref{dp-bd}(b1)). The $\mathbb{L}_{2}$-norm error between learning solution $\widehat{u}(x,t)$  and numerical solution is 2.74e-02.

\begin{figure}[t]
\begin{center}
\vspace{-0.06in}
\hspace{0.2in} {\scalebox{0.75}[0.75]{\includegraphics{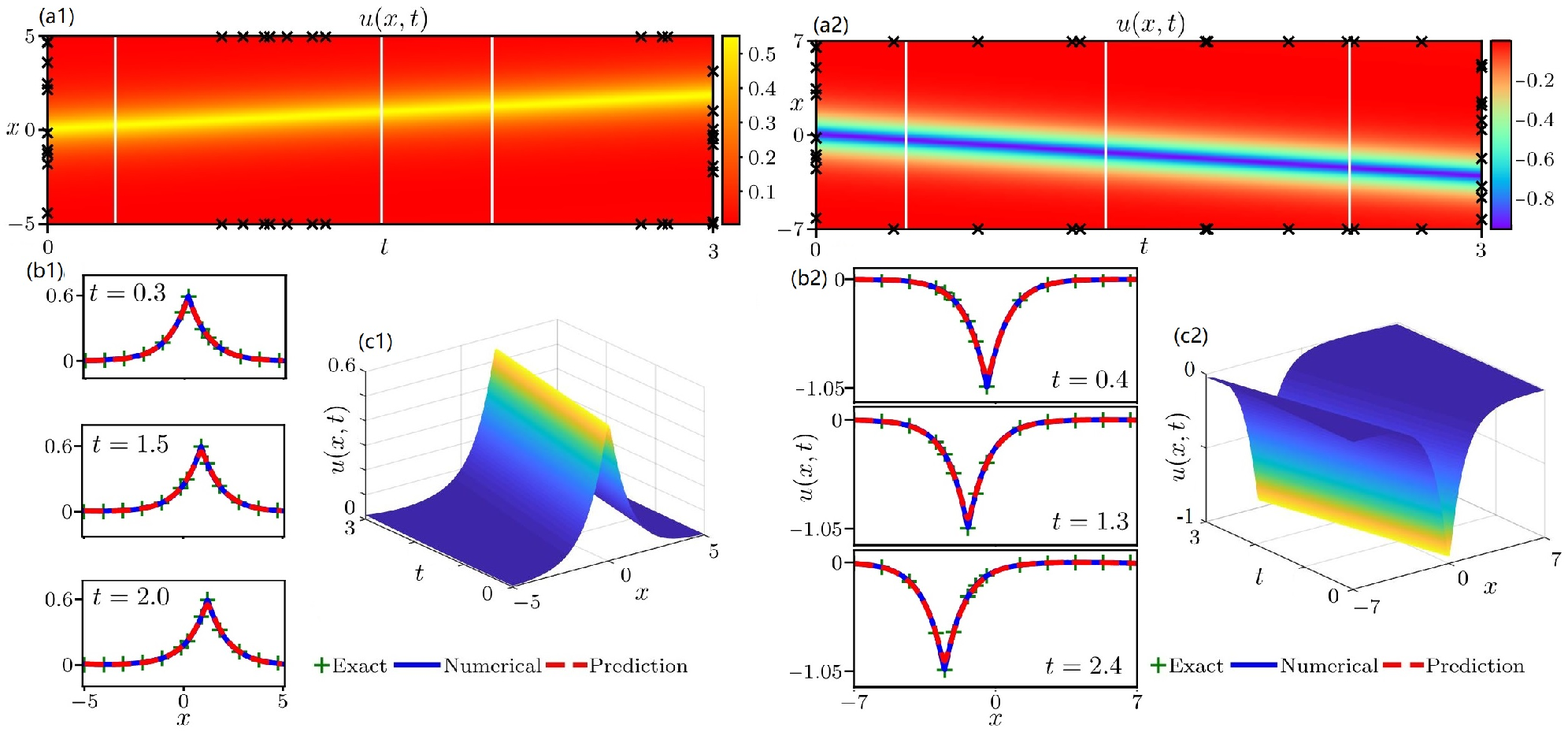}}}
\end{center}
\par
\vspace{-0.06in}
\caption{{\protect\small The DP equation. (a1,a2) The data-driven peakon solutions resulted from the PINN; (b1,b2) The comparisons between the learning, numerical, and exact peakon solutions at the distinct times (a1) $t = 0.3,\, 1.5$, and $2$, (a2) $t = 0.4,\, 1.3$, and $2.4$; The $\mathbb{L}_{2}$-norm errors $\widehat u(x,t)$ between learning and numerical peakon solutions are (b1) 2.74e-02 and (b2) 3.29e-02; (c1,c2) The 3D profiles of the learning bright and dark peakon solutions. $c=0.6$ for (a1, b1, c1) and $c=-1.05$ for (a2, b2, c2).}}
\label{dp-bd}
\end{figure}
\begin{figure}[ht]
\begin{center}
\vspace{-0.1in} {\scalebox{0.45}[0.45]{\includegraphics{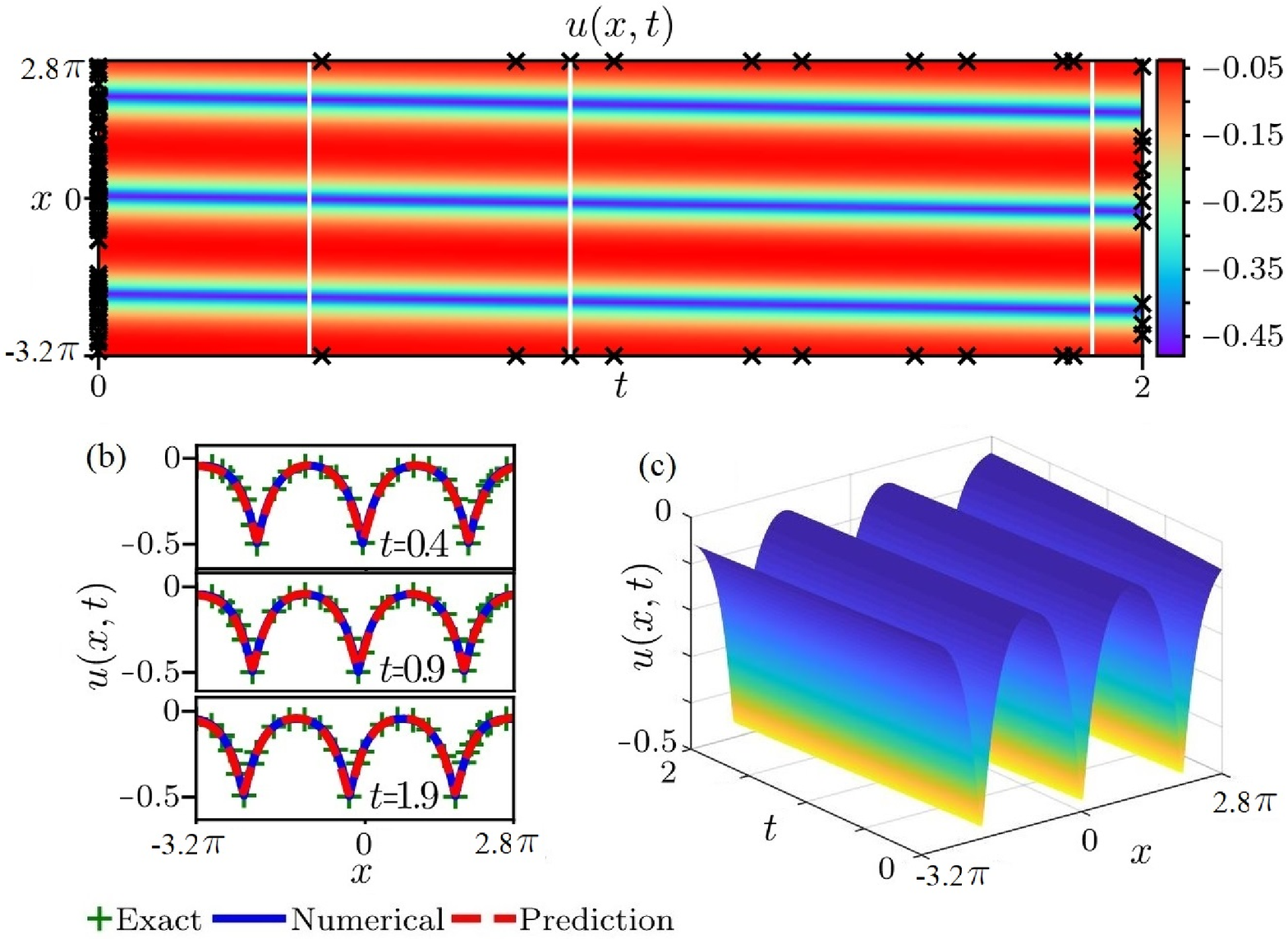}}}
\end{center}
\par
\vspace{-0.15in}
\caption{{\protect\small The DP equation. (a) The data-driven periodic peakon solution resulted from the PINN, and three distinct tested times; (b) The comparisons between the learning, numerical, and exact periodic peakon solutions
at the distinct times $t = 0.4$, $0.9$, and $1.9$. The $\mathbb{L}_{2}$-norm error  between the learning and numerical periodic peakon solutions is  1.96e-02; (c) The 3D profile of the learning periodic peakon solution.}}
\label{fdp-p}
\end{figure}

 ii) For the case $c=-1.05$ in the initial condition (\ref{ch-int}), we consider the spatio-temporal region $(x,t)\in [-7,7]\times [0, 3]$ with 256 Fourier modes and time-step $\Delta t = 0.006$. The training data-set used in the 5-hidden-layer deep PINN $f_{dp}(x,t)$ with 20 neurons per layer consists of randomly chosen $N_{int}=10$ points from the initial data $u(x,0)$, $N_{end}=10$ points from the end data $u(x,3)$,  $N_b=10$ points pertaining to the periodic boundary data, and $N_f=1000$ points from the spatial-temporal solution zone.
Figures~\ref{dp-bd}(a2) and (c2) exhibit the 2D and 3D profiles of the latent solution $\widehat u (x,t)$.
 The comparison of learning solution (red dashed line), numerical solution (blue solid line) and exact solution (green plus)
 is shown at three different times $t = 0.4, 1.3$ and $2.4$ (see Fig.~\ref{dp-bd}(b2)). The $\mathbb{L}_{2}$-norm error between learning solution $\widehat{u}(x,t)$  and numerical solution is 3.29e-02.

\v {\it Case 2.} Since the DP equation (\ref{dp}) admit the same periodic peakon solution as one of the CH equation, thus we consider
 the initial value condition (\ref{ch-p}) with $c=-0.5$ and periodic boundary conditions. Similarly, The considered spatio-temporal region in the  pseudo-spectral method is $(x,t)\in [-16\pi/5, 14\pi/5]\times [0, 2]$ with 512 Fourier modes and time-step $\Delta t = 0.004$.
The training data-set used in the 6-hidden-layer deep PINN $f_{dp}(x,t)$ with 20 neurons per layer consists of randomly chosen $N_{int}=90$ points from the initial data $u(x,0)$, $N_{end}=10$ points from the end data $u(x,2)$,  $N_b=10$ points pertaining to the periodic boundary data, and $N_f=1000$ points from the spatial-temporal solution zone by minimizing the MSE loss (\ref{mse}).

Figures~\ref{fdp-p}(a) and (c) exhibit the 2D and 3D profiles of the latent periodic peakon solution $\widehat u (x,t)$. The comparison of learning solution (red dashed line), numerical solution (with blue solid line) and exact solution (green plus) is shown at three different times $t = 0.4,\, 0.9$, and $1.9$ (see Fig.~\ref{fdp-p}(b)).  The $\mathbb{L}_{2}$-norm error between learning solution $\widehat{u}(x,t)$  and numerical solution is 1.96e-02.

\end{document}